\documentclass[prx,amsmath,amssymb,twocolumn,floatfix]{revtex4-2}
\usepackage{graphicx}
\usepackage{natbib}
\usepackage[utf8]{inputenc}
\usepackage[squaren,Gray]{SIunits}
\usepackage{physics}
\newcommand{\KK}{\mathrm{KK}}
\begin{document}

\title{Quantum bath engineering of a high impedance microwave mode through quasiparticle tunneling}

\author{Gianluca Aiello$^1$}
\affiliation{$^{1}$Laboratoire de Physique des Solides, CNRS, Université Paris Saclay, Orsay, France}
\author{Mathieu Féchant$^1$}
\affiliation{$^{1}$Laboratoire de Physique des Solides, CNRS, Université Paris Saclay, Orsay, France}
\author{Alexis Morvan$^1$}
\affiliation{$^{1}$Laboratoire de Physique des Solides, CNRS, Université Paris Saclay, Orsay, France}
\author{Julien Basset$^1$}
\affiliation{$^{1}$Laboratoire de Physique des Solides, CNRS, Université Paris Saclay, Orsay, France}
\author{Marco Aprili$^1$}
\affiliation{$^{1}$Laboratoire de Physique des Solides, CNRS, Université Paris Saclay, Orsay, France}
\author{Julien Gabelli$^1$}
\affiliation{$^{1}$Laboratoire de Physique des Solides, CNRS, Université Paris Saclay, Orsay, France}
\author{Jérôme Estève$^1$}
\affiliation{$^{1}$Laboratoire de Physique des Solides, CNRS, Université Paris Saclay, Orsay, France}

\begin{abstract}
We demonstrate a new approach to dissipation engineering in microwave quantum optics. In this context, dissipation usually corresponds to quantum jumps, where photons are lost one by one. 
By coupling a high impedance microwave resonator to a tunnel junction, we use the photoassisted tunneling of quasiparticles as a tunable dissipative process. 
We are able to adjust the minimum number of lost photons per tunneling event to be one, two or more, through a dc voltage.
Consequently, different Fock states of the resonator experience different loss processes.
Causality then implies that each state experiences a different energy (Lamb) shift, as confirmed experimentally.
This photoassisted tunneling process is analogous to a photoelectric effect, which, for the first time, requires a quantum description of light to be quantitatively understood.
This work opens up new possibilities for quantum state manipulation in superconducting circuits, which do not rely on the Josephson effect.
\end{abstract}

\maketitle

Quantum bath engineering is considered as a promising route to perform certain tasks in quantum information processing, such as state stabilization, passive error correction or fast qubit initialization \citep{sarovar_continuous_2005,kraus_preparation_2008,verstraete_quantum_2009,reed_fast_2010,krauter_entanglement_2011,barreiro_open-system_2011,murch_cavity-assisted_2012,leghtas_confining_2015}.
In the context of circuit QED, bath engineering usually results from the interplay between coherent evolution and dissipation in the form of single photon loss \citep{kapit_upside_2017}.
Such engineered losses, in particular two photon losses, are at the heart of promising error correction schemes in superconducting qubit architectures \citep{leghtas_confining_2015}.   
Here, we demonstrate a different approach where engineered dissipation comes from the non-linear coupling of a microwave mode to a tunnel junction, which realizes a bath consisting of two electronic reservoirs.
Dissipation arises from the photoassisted tunneling processes, during which one electron tunnels, while $l$ photons are absorbed from the mode.  
Because the mode is sustained by a high kinetic inductance superconducting resonator made of granular Aluminum, its characteristic impedance is sufficiently large such that the high order loss processes with $l>1$ are allowed \citep{catelani_relaxation_2011,silveri_theory_2017,esteve_quantum_2018,silveri_broadband_2019,viitanen_photon-number-dependent_2021,vadimov_single-junction_2022}.
The rate of processes with given $l$ can be tuned through the dc voltage that biases the junction. 
As an example of engineered dissipation, we focus on the regime where $l \geq 2$ processes dominate over single photon loss.
The dynamics is then restricted by the quantum Zeno effect to the subspace spanned by the zero and one photon Fock states  \citep{facchi_quantum_2008}, turning the harmonic oscillator mode into a two-level system.

From a broader perspective, photoassisted tunneling is a special case of photoelectric effect, where the electron is emitted into a contact rather than in free space.
As with the photoelectric effect, the frequency of the light must exceed the chemical potential difference between the two contacts divided by the Planck constant in order to observe photoassisted tunneling at low light intensity.
The natural interpretation for this threshold behavior uses the concept of photon as discussed above.
But, in most cases, the electric field that is responsible for the electron emission may be considered as a classical field.
Even though the photoelectric effect lead Einstein to propose the idea of photon, the standard semiclassical model of light matter interaction, which neglects the quantum nature of light and treats only matter at the quantum level, accurately describes all photoemission experiments \cite{reinert_photoemission_2005}.
This paradox has been known and debated for a long time \cite{scully_concept_2008}.
Our experiment sheds new light on this problem by reaching a regime, where both matter and light must be treated at the quantum level in order to reach a quantitative understanding.
In the context of a microwave resonator coupled to a tunnel junction, the semiclassical approach describes the junction in terms of an admittance, which can then be used to model its effect on the resonator mode coupled to the junction \citep{tucker_quantum_1985,worsham_quantum_1991}.
In the last part of the paper, we compare the predictions of this model to the ones of the full quantum model, in particular for the frequency shift of the resonator. 
Our data confirm that quantum effects significantly contribute to the induced energy shift, the so-called Lamb shift \citep{breuer_theory_2007}. Furthermore, the energy shift is different for each Fock state, which is meaningless in a classical model.

\begin{figure}
    \includegraphics{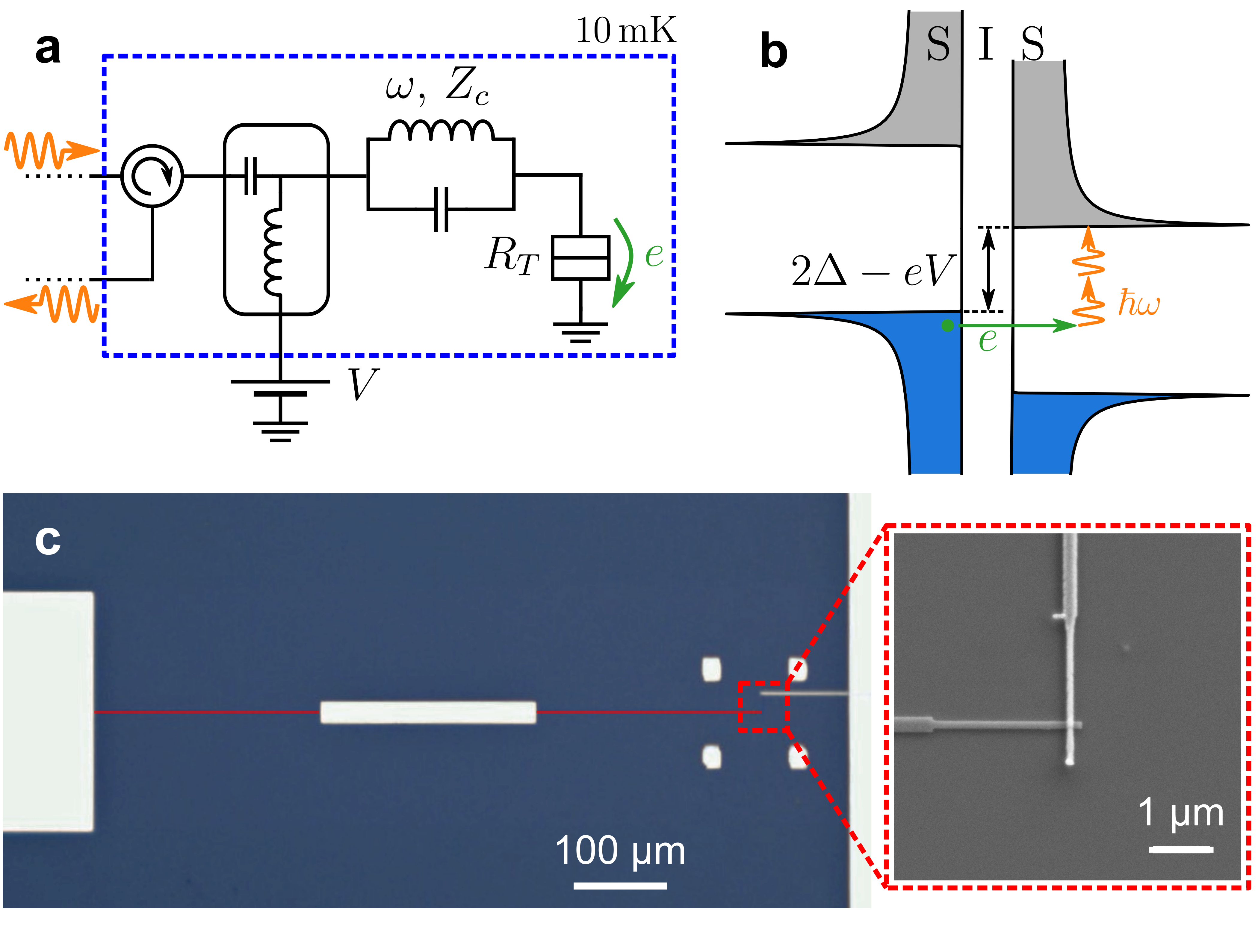}
    \centering
    \caption{{\bf Experiment principles.} {\bf a} Schematic of the experimental circuit. 
    A microwave mode at frequency $\omega \approx 2\pi \times \unit{6}{\giga\hertz}$, here represented by a $LC$ resonator, is coupled to a superconducting tunnel junction with tunnel resistance $R_T$.
    The characteristic impedance $Z_c$ of the mode is \unit{4.5}{\kilo\ohm}, much larger than in a conventional superconducting resonator.
    A bias tee and a circulator are used to dc bias the sample while measuring the reflected microwave signal.
    {\bf b} When  the bias voltage is such that $2\Delta - eV > l\hbar \omega$, the tunneling of quasiparticles through the junction is allowed only if at least $l$ photons (here two) are absorbed from the mode to provide the missing energy.
    {\bf c} Microscope image of the sample realizing the circuit shown in {\bf a}.
    The resonator consists of two grAl quarter wave resonators (red) connected by a wider Al wire (white).
    The tunnel junction (Al/AlO$_x$/Al) connects the resonator right end to the ground.
    The junction area is $150 \times 150 \, \mathrm{nm}^2$, leading to a tunnel resistance $R_T = \unit{150}{\kilo\ohm}$ far above the gap.
    A microstrip Al line connects the resonator left end to the measurement circuit. All experiments are performed in a dilution fridge with a base temperature of \unit{10}{\milli\kelvin}.}
\end{figure}

The principle of the experiment is presented in figure~1. A high impedance resonator with a resonant mode around $\omega \approx 2\pi \times \unit{6}{\giga\hertz}$ is galvanically coupled to a tunnel junction as schematically depicted in figure~1a.
In addition to the usual single photon loss, due to the coupling to the measurement line or to intrinsic loss mechanisms, photons in the mode may also be absorbed through the photoassisted tunneling of a quasiparticle across the junction (figure 1b) \citep{tien_multiphoton_1963,tucker_quantum_1985}.
Such inelastic tunneling processes, where $l$ photons are absorbed, are energetically allowed only if the bias voltage $V$ is sufficiently close to the gap, $e V > 2 \Delta - l \hbar \omega$, where $\Delta \approx \unit{200}{\micro\electronvolt}$ is the superconducting gap in each Al electrode.
The junction thus realizes a tunable quantum absorber, where the minimum number of absorbed photons per photoassisted tunneling event is set by the voltage bias. 

In order for the engineered loss to be efficient, the corresponding rate must dominate single photon loss, which is only possible if the characteristic impedance of the mode coupled to the junction is sufficiently large.  
The expected rate for the photoassisted tunneling process can be derived from the tunnel Hamiltonian, which describes the coupling between the tunneling electrons and an electromagnetic mode \citep{devoret_effect_1990}
\begin{equation}
    \hat{H}_{\mathrm T} = e^{i\lambda (\hat{a}+\hat{a}^\dagger)} \hat{B} +{\rm h.c.} \, . \label{eq.tunnel}
\end{equation}
The operator $\hat{a}$ is the annihilation operator of the considered mode and $\hat{B}=T \sum_{LR} \hat{c}^\dagger_R \hat{c}_L$ transfers one electron from the left ($L$) to the right ($R$) junction contact.
The barrier transparency $T$ is inversely proportional to the junction resistance $R_T$ at voltages far above the gap.
The displacement operator $e^{i\lambda (\hat{a}+\hat{a}^\dagger)}$ can be interpreted as a consequence of charge conservation: one tunneling event corresponds to a translation of the charge degree of freedom by one electron \citep{devoret_effect_1990}.
The displacement amplitude, $\lambda$, is proportional to the zero point fluctuations of the conjugate of the charge operator and is given by $\lambda = \sqrt{\pi Z_c/R_K}$, where $Z_c$ is the characteristic impedance of the mode and $R_K=h/e^2$ is the quantum of resistance.
This is similar to the displacement operator that appears in the coupling Hamiltonian between light and an atom trapped in an harmonic potential as a consequence of momentum conservation \citep{meekhof_generation_1996}. 

The matrix elements of the displacement operator between two Fock states $\ket{n}$ and $\ket{n+l}$ with $l \geq 0$ are the Franck-Condon coefficients, which depend on $\lambda$ as \citep{catelani_relaxation_2011,souquet_photon-assisted_2014,qassemi_quantum_2015}
\begin{equation*}
    \alpha_{nl} = | \mel{n+l}{e^{i\lambda (\hat{a}+\hat{a}^\dagger)}}{n} |^2 = \frac{  \lambda^{2l} e^{-\lambda^2} n!}{(n+l)!}  L_n^{(l)}(\lambda^2)^2 \, ,
\end{equation*}
where $L_n^{(l)}$ are the generalized Laguerre polynomials.
In a standard superconducting resonator, $Z_c$ is much smaller than $R_K$, resulting in $\lambda \ll 1$, which is analogue to the Lamb-Dicke regime for atoms \citep{meekhof_generation_1996}. 
In this case, processes between Fock states differing by $l$ are exponentially suppressed as $\lambda^{2l}$.  
Here, we are interested in the opposite regime, where $\lambda \sim 1$. In this case, the displacement amplitude in the mode quadrature phase space is comparable to the extension of the ground state and transitions between different Fock states are allowed.

Considering the left and right contacts as electronic reservoirs at thermal equilibrium, the Fermi golden rule predicts that photoassisted tunneling leads to a loss rate for the $\ket{n}$ state given by \citep{catelani_relaxation_2011,souquet_photon-assisted_2014,mendes_cavity_2015,silveri_theory_2017,esteve_quantum_2018}
\begin{equation}
    \gamma_n = \frac{1}{e} \sum_{l=1}^n \alpha_{n-l,l} I(V + l\hbar \omega /e ) \, , \label{eq.gamma}
\end{equation} 
where each term in the sum corresponds to the contribution of the $l$ photon absorption process.
The rate of such process is proportional to the corresponding Franck-Condon factor multiplied by the current $I$ that would flow through the junction in the absence of resonator.
Energy conservation implies that the current must be evaluated at the voltage corresponding to the bias voltage shifted by the energy of the $l$ photons.
Because of the superconducting gap, the $l$ photon process is allowed only when $eV + l\hbar \omega  \geq 2\Delta$, otherwise $I=0$. 
Note that $I(V)$ coincides with the actual current flowing through the junction only when photoassisted processes are negligible, \emph{i.e.} at large voltages above the gap.

In the experiment presented here, we reach $Z_c = \unit{4.5}{\kilo\ohm}$, which corresponds to $\lambda=0.74$. 
In this regime, high order processes have Franck-Condon coefficients that are comparable to the one of the one photon processes, e.g. $\alpha_{02} \approx \alpha_{01}/3 \approx 0.1$.
Figure 2a shows the evolution of $\gamma_n$ as a function of voltage for parameters corresponding to our experiment.
The $I(V)$ characteristic of the junction is calculated from the resistance $R_T=\unit{150}{\kilo\ohm}$ measured far above the gap.
But, in order to take into account the presence of other modes in the resonator, we replace $R_T$ by an effective tunnel resistance with a larger value $\tilde{R}_T=\unit{430}{\kilo\ohm}$.
The increase of resistance is given by the product of the dynamical Coulomb blockade factors $\Pi_{n \neq 1} e^{\lambda_n^2}$ over all the modes except the one at \unit{6}{\giga\hertz} (see SI).
The figure 2a shows that the junction is expected to act as a tunable absorber that can distinguish between the first Fock states up to $n=3$.  

In order to reach $\lambda \sim 1$, we use granular Aluminum (grAl) as the material of the resonator.
GrAl has been shown to be a promising material to realize a superinductance with small loss\cite{Grunhaupt2018, Maleeva2018, Kamenov2020,glezer_moshe_granular_2020}. 
Other possible methods include resonators with carefully designed geometries  \citep{rolland_antibunched_2019,peruzzo_surpassing_2020}, the use of other high kinetic inductance superconductors \citep{annunziata_tunable_2010, Barends2010, Samkharadze2016} or chains of Josephson junctions \citep{manucharyan_fluxonium_2009,bell_quantum_2012,masluk_microwave_2012,leger_observation_2019}. 
To first approximation, the mode probed in the experiment is the fundamental mode of a quarter wavelength resonator made of a \unit{0.5}{\micro\meter} wide and \unit{200}{\micro\meter} long grAl  wire with a kinetic inductance of $\unit{0.56}{\nano\henry}/\Box$ (see figure 1c). 
The junction connects the end of the resonator, where the mode has a voltage maximum, to the ground.
In order to obtain a high quality factor, despite the galvanic connection to the measurement line, a distributed Bragg reflector (DBR) is inserted between the resonator and the line.

The resonant frequencies, characteristic impedances and coupling loss rates of the different modes sustained by the structure are numerically simulated (see SI). 
The design is chosen to obtain a mode with a large characteristic impedance, a resonance frequency close to \unit{6}{\giga\hertz} and a quality factor above $10^4$. 
The sample is fabricated through standard e-beam lithography and double angle evaporation (see SI).
The exact value of $\lambda$ depends on the precise value of the kinetic inductance of the wire and the junction capacitance, which can only be estimated at the design stage.
We deduce the precise values of these two parameters by comparing the simulation to the measured resonance frequencies of the \unit{6}{\giga\hertz} mode as well as the ones of the other modes at 1.9, 12, 24 and \unit{32}{\giga\hertz}.
We finally obtain $Z_c = \unit{4.5}{\kilo\ohm}$ for the \unit{6}{\giga\hertz} mode and a quality factor of $1.3 \times 10^4$, which corresponds to a coupling loss rate to the measurement line $\kappa_c = 2\pi \times \unit{0.45}{\mega\hertz}$.

\begin{figure}
    \centering
    \includegraphics{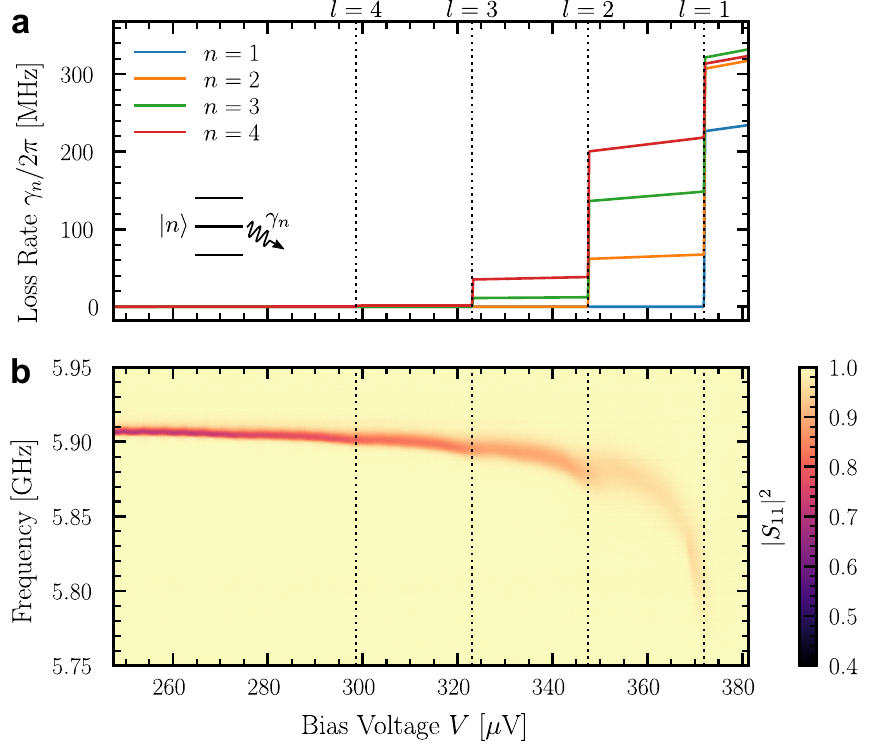}
    \caption{{\bf Tunnel junction as a tunable quantum absorber.} {\bf a}  Quasiparticle tunneling induces an extra loss $\gamma_n$ (eq. \ref{eq.gamma}), which is different for each Fock state of the mode coupled to the junction. 
    Dashed vertical lines show the onset of the $l$ photon absorption process, given by $(2\Delta-l\hbar \omega)/e$.
    The photon number $n$ corresponding to the lowest Fock state with increased loss can be chosen with the bias voltage.
    The parameters correspond to the ones expected in the experiment.
    {\bf b} Experimental spectroscopy of the \unit{6}{\giga\hertz} mode as a function of the bias voltage $V$. 
    The image shows the measured reflection coefficient $|S_{11}|^2$ for an incident power of -115\,dBm on the resonator.
    The onset of the different absorption processes are clearly visible up to $l=4$.}
\end{figure}

Figure 2b shows the spectroscopy of the \unit{6}{\giga\hertz} mode as a function of the voltage bias close to the superconducting gap $2\Delta/e$. 
The incoming microwave power is chosen to populate many Fock states so that we can observe the onset of the different loss process every time $eV > 2\Delta -  l\hbar \omega$ (vertical dashed lines).
When a new loss process is allowed, the intensity in the mode decreases, leading to a diminution of the reflection dip.
At the same time, the width of the resonance increases.
When the one photon absorption process becomes allowed, the resonance abruptly disappears.
This is because the corresponding loss rate, which affects every Fock states, is much larger than $\kappa_c$ (see figure 2a).
The mode becomes under-coupled and the reflection dip vanishes.
At the same time that loss increases as the voltage increases, the resonance frequency redshifts as a consequence of the Kramers-Kronig relations. 
In particular, we observe small frequency kinks every time a new loss process appears. 
These frequency shifts will be detailed at the end of the manuscript. At lower voltages (not shown here), we observe multiple kinks in the spectrum that correspond to inelastic Cooper pair tunneling resonances. These results and their analysis will be presented elsewhere.

\begin{figure}
    \centering
    \includegraphics{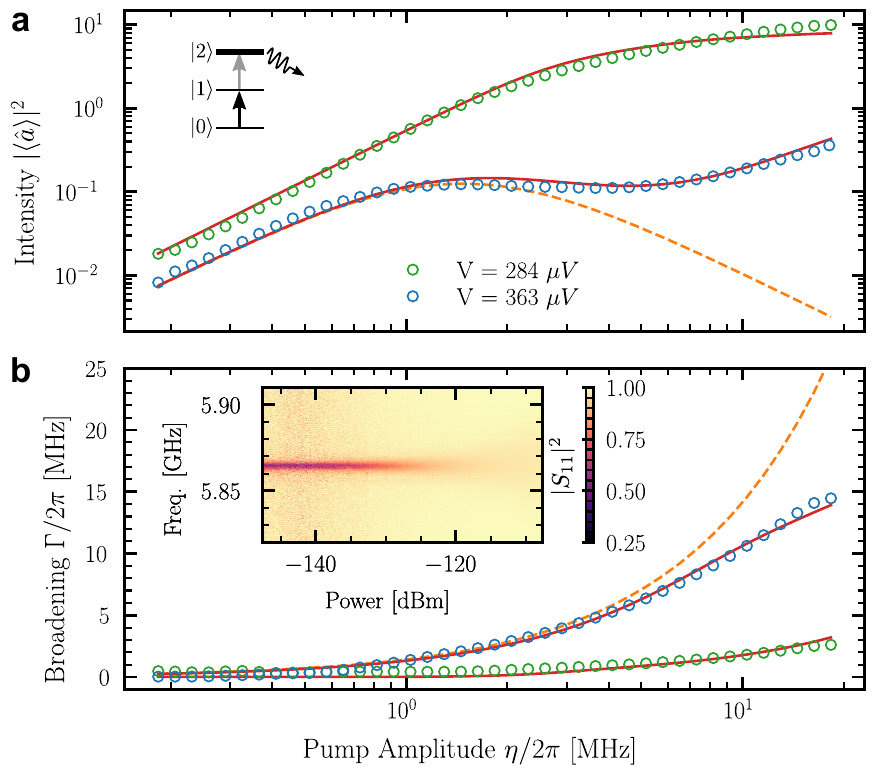}
    \caption{{\bf Quantum Zeno dynamics.} {\bf a} Evolution of the squared mean amplitude $|\langle a\rangle|^2$ in the mode as a function of the pump amplitude $\eta$ for two different bias voltages. In the absence of Zeno effect, $|\langle a\rangle|^2$ is quadratic with the pump amplitude (green data). 
    When the voltage lies in the range where the quantum Zeno dynamics limits the dynamic to the one of a two level system, we observe a clear saturation (blue data).
    The solid red line shows the prediction of a master equation taking into account the different absorption rates for different Fock states (see SI).
    The dashed line shows the expectation for an ideal two level system.
    The calibration of the measured intensity is detailed in the SI.
    {\bf b} Evolution of the power broadening $\Gamma$ as a function of the pump amplitude. The resonance spectra shown in the inset are fitted using the usual formula for a two-level system, which predicts a fwhm $\sqrt{\kappa^2+2\Gamma^2}$ (see SI), where $\kappa$ is the total loss rate. 
    Power broadening is important in the case of Zeno dynamics (blue data) and negligible otherwise, except at very high power (green data).
    The solid and dashed lines show the result of the master equation simulation and the expectation for an ideal two level system.
    }
\end{figure}

The step like increase of the $l$ photon loss rate at $eV = 2\Delta - l\hbar \omega$ is ideally suited to induce quantum Zeno dynamics and engineer the Hilbert space of the \unit{6}{\giga\hertz} mode.
If the voltage is set in the range $2\Delta - \hbar \omega > eV > 2\Delta - 2\hbar \omega$, all Fock states except $\ket{0}$ and $\ket{1}$ are lossy because of $l\geq2$ processes. 
In particular, the $\ket{2}$ state experiences a two photon loss rate on the order of $\gamma_2 \approx 2\pi \times \unit{65}{\mega\hertz}$, while the single photon loss induced by the junction is expected to be negligible.
We bias the junction in this voltage range, pump the mode with a microwave tone and measure the reflected signal with an homodyne detector.
The intensity in the mode at resonance is shown in figure 3a as a function of the pump amplitude $\eta$, which is related to the incoming pump power $P$ as $\eta = \sqrt{\kappa_c P /\hbar \omega}$.
Because of the engineered dissipation, a state initially in the subspace spanned by $\ket{0}$ and $\ket{1}$ is continuously projected in this subspace in the absence of tunneling event.
This non destructive measurement induces quantum Zeno dynamics in this subspace and the mode behaves as a two-level system rather than an harmonic oscillator.
This restriction of the Hilbert space is efficient as long as $\eta$ remains small compared to the projection rate, which is here set by the photon loss rate $\gamma_2$.
When this is the case, we observe a saturation of the intensity near $|\langle a\rangle|^2=1/8$, as expected for a two level system.
This is a clear signature of the reduction of the Hilbert space to the $\ket{0}$ and $\ket{1}$ subspace.
At high pump power, the quantum Zeno effect breaks down and the mode intensity starts to increase again even though more and more photons are absorbed by the junction. 
In the two-level saturation regime, we also observe that the resonance width increases because of power broadening (see figure 3b).
We use this effect in order to calibrate the pump amplitude by assuming that the broadening is linear in pump intensity at low pump power, as with an ideal two-level system.
The solid lines in figure 3 show the results of the numerical simulation of a master equation describing the evolution of the mode coupled to the junction as detailed in the SI \citep{silveri_theory_2017,esteve_quantum_2018}.
Our data are well reproduced by a single mode model using the expected effective tunnel resistance $\tilde{R}_T$ and including an additional single photon loss rate of a few MHz. 
The simulation includes the Lamb shift of the different levels, which also contributes to the blockade of the $1 \rightarrow 2$ transition (see below). 
The figure of merit of the observed Zeno blockade can be quantified by the ratio between the effective loss rate from $\ket{2}$ and from $\ket{1}$ that reproduces our data, which is around 25 for our experiment.

\begin{figure}
    \centering
    \includegraphics{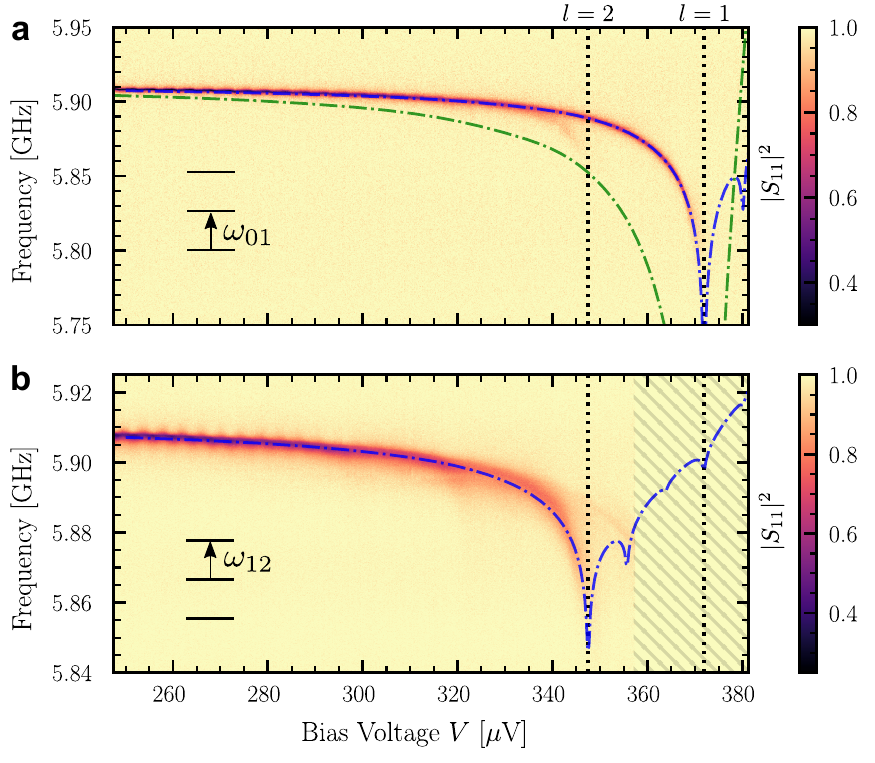}
    \caption{{\bf One and two photon spectroscopy as a function of voltage.}
    {\bf a} Reflected signal measured with an injected microwave power of \unit{-140}{dBm} as a function of frequency and bias voltage. The power is sufficiently low to mostly probe the $0\rightarrow 1$ transition. The vertical lines are the same than in figure 1.
    {\bf b} Reflected signal in the presence of a second microwave tone tuned at the frequency measured in {\bf a}. 
    This two photon spectroscopy probes the $1\rightarrow 2$ transition, which shifts differently than the $0\rightarrow 1$ transition, showing the non-linearity induced by the Lamb shift.
    In both cases, the dashed blue lines correspond to the prediction of an \emph{ab initio} quantum model. The dashed green line in {\bf a} corresponds to the classical approximation, keeping only terms of order $\lambda^2$ in the expression of the frequency shift. The hatched area was not measured. 
    }
\end{figure}

We now turn to a detailed analysis of the shift of the resonance frequency as a function of the voltage bias.
The frequency shift experienced by each Fock state is related to the loss rate via the Kramers-Kronig relations.
We introduce the Kramers-Kronig (KK) transform of the current voltage characteristic as \citep{tucker_quantum_1985}
\begin{equation*}
    I^{\KK}(V) = \frac{1}{\pi} \mathcal{P} \int_{-\infty}^{\infty} \frac{I(V')}{V'-V} \, dV' \, .    
\end{equation*}
The frequency shift of the $\ket{n}$ state, which is usually called Lamb-shift \citep{breuer_theory_2007}, is then given by \citep{esteve_quantum_2018}
\begin{equation}
    \delta \omega_n = -\frac{1}{2e} \left( \sum_{l=1}^n \alpha_{n-l,l} I^{\KK}_l  +  \sum_{l=0}^\infty \alpha_{n,l} I^{\KK}_{-l} \right) \, , \label{eq.wn}
\end{equation}
where $I^{\KK}_l$ stands for $I^{\KK}(V+l\hbar \omega/e)$. 
The first term corresponds to the KK transform of (\ref{eq.gamma}). 
The second term is absent from (\ref{eq.gamma}), because the corresponding rates vanish for voltages below the gap, which is not the case for the KK transform, \emph{i.e} $I^{\KK}(V)$ is not zero when $eV<2\Delta$ (see SI for a plot of $I^{\KK}(V)$).
From (\ref{eq.wn}), the shift of the resonance frequency for the fundamental $0\rightarrow 1$ transition is given by 
\begin{equation}
    \delta \omega_{01} = -\frac{\lambda^2 e^{-\lambda^2}}{2e} \sum_{l=0}^\infty \frac{\lambda^{2l}}{l!}(I^{\KK}_{-l-1} + I^{\KK}_{-l+1} - 2 I^{\KK}_{-l}) \label{eq:expansion}
\end{equation}
Keeping only the $l=0$ term and neglecting the $e^{-\lambda^2}$ term leads to the frequency shift that is derived from a classical treatment of the electromagnetic field \citep{tucker_quantum_1985}, in which case the reactive part of the junction admittance at low intensity can be written as $e(I^{\KK}_1 + I^{\KK}_{-1} - 2 I^{\KK}_{0})/(2\hbar \omega)$ as measured in \citep{worsham_quantum_1991,basset_high-frequency_2012,silveri_broadband_2019}.
This classical expression also corresponds to the first term of the Taylor expansion of (\ref{eq:expansion}) in powers of $\lambda$.
In our case, because $\lambda$ is large, this approximation is not valid and the higher order terms are expected to significantly contribute to the frequency shift. 

Figure 4 shows the reflection spectrum that we measure at very low pump power in order to probe the $0 \rightarrow 1$ transition only. 
We compare it to the predictions of the classical admittance model (dashed green line) and to the one of the quantum model (dashed blue line). 
Because of the presence of other modes in the resonator than the \unit{6}{\giga\hertz} mode, the expression for $\delta \omega_{n}$ that we use in the quantum description is slightly more involved than the one given in (\ref{eq.wn}) and is given in the SI. 
As expected, the two models significantly differ and our data are in good agreement with the \emph{ab initio} quantum model. 
This is a rare situation, similar to the original Lamb shift effect \citep{maclay_history_2020}, where quantum effects significantly affect the frequency shift.
The $\lambda$ coefficient may be rewritten as $\lambda^2= 2\pi \alpha Z_c/Z_{\rm vac}$, where $\alpha\approx 1/137$ is the fine structure constant and $Z_{\rm vac}\approx \unit{377}{\ohm}$ is the impedance of free space, showing that the expansion (\ref{eq:expansion}) in powers of $\lambda$ is actually an expansion in powers of $\alpha$ as expected for a QED effect.
A similar result was obtained in the dual situation where a transmon qubit is frequency shifted by a high impedance environment \citep{leger_observation_2019}. 
Equation (\ref{eq:expansion}) shows that the quantum correction are more than a simple renormalization of the resistance by the factor $e^{\lambda^2}$. 

Equation (\ref{eq.wn}) also predicts that the Lamb shift terms introduces a non-linearity in the harmonic spectrum of the mode as a consequence of the non-linear bath coupling. 
This effect is already visible in figures 2b and 4a, where we observe a shift and even a splitting of the resonance when $eV \approx 2 \Delta - l \hbar \omega$. 
In order to confirm that this splitting can be attributed to different shifts of the different Fock states, we perform a two photon spectroscopy as shown in figure~4b. 
A first tone is tuned to excite the $0 \rightarrow 1$ resonance that we measure at very low power (fig.~4a).
We then acquire a reflection spectrum using a second tone that mostly probes the $1 \rightarrow 2$ transition. 
The signal is only visible when $eV \le  2 \Delta - 2 \hbar \omega$ for the same reason as in fig.~1a. 
We clearly observe a frequency shift of the $1 \rightarrow 2$ transition, which is different from the one of the $0 \rightarrow 1$ transition, in very good agreement with the quantum model. 
This non-linear effect favors the observed restriction of the Hilbert space to the first two levels. 
The induced non-linearity is maximum around \unit{383}{\micro \volt} and equal to $(\omega_{12} - \omega_{01})/2\pi \simeq \unit{42}{\mega \hertz}$.
The non-linear shifts due to the $l>2$ terms in (\ref{eq.wn}) are responsible of the kinks in the resonance frequency at $eV=2\Delta - l \hbar \omega$ observed in figure 2b. 

In conclusion, we have demonstrated a new way to engineer dissipation in superconducting QED circuits by taking advantage of the non-linear coupling between a high impedance mode and electronic reservoirs.
The dominant loss mechanism can be tuned to be a one, two, or even higher order photon process. 
Our results could be extended to other types of junctions with a non-linear current voltage characteristic. 
Such engineered dissipation could have applications in quantum computing for rapid initialization of a microwave mode to vacuum, or to stabilize states in error correction schemes.
More fundamentally, our results give an example of a situation where quantum effects invalidate the classical approach to dissipation based on linear response theory, for example in terms of admittance, to describe the coupling between the different elements of a circuit.

The authors would like to thank Claire Marrache-Kikuchi and Hélène Le Sueur for their collaboration at early stages of the experiment and Richard Deblock for fruitful discussions. This work is supported by the Agence Nationale de la Recherche (ANR-18-CE47-0003 BOCA project) and the Laboratoire d’excellence Physique Atomes Lumière Matière (ANR-10-LABX-0039-PALM).

\bibliography{biblio}

\end{document}


\renewcommand{\theequation}{S.\arabic{equation}}
\renewcommand{\thetable}{S\arabic{table}}
\renewcommand{\thefigure}{S\arabic{figure}}

\title{Quantum bath engineering of a high impedance microwave mode through quasiparticle tunneling: Supplementary Information}

\author{Gianluca Aiello$^1$}
\affiliation{$^{1}$Laboratoire de Physique des Solides, CNRS, Université Paris Saclay, Orsay, France}
\author{Mathieu Féchant$^1$}
\affiliation{$^{1}$Laboratoire de Physique des Solides, CNRS, Université Paris Saclay, Orsay, France}
\author{Alexis Morvan$^1$}
\affiliation{$^{1}$Laboratoire de Physique des Solides, CNRS, Université Paris Saclay, Orsay, France}
\author{Julien Basset$^1$}
\affiliation{$^{1}$Laboratoire de Physique des Solides, CNRS, Université Paris Saclay, Orsay, France}
\author{Marco Aprili$^1$}
\affiliation{$^{1}$Laboratoire de Physique des Solides, CNRS, Université Paris Saclay, Orsay, France}
\author{Julien Gabelli$^1$}
\affiliation{$^{1}$Laboratoire de Physique des Solides, CNRS, Université Paris Saclay, Orsay, France}
\author{Jérôme Estève$^1$}
\affiliation{$^{1}$Laboratoire de Physique des Solides, CNRS, Université Paris Saclay, Orsay, France}

\maketitle

\section{Sample geometry and EM simulations}\label{sec:emsimulation}
The resonator consists in a quarter wavelength microstrip line, which is galvanically coupled to a \unit{50}{\ohm} line through a distributed Bragg reflector (DBR). In order to obtain high impedance modes, the resonator is made of granular Aluminum (grAl), which has a high kinetic inductance, and the width of the microstrip line is reduced to \unit{0.5}{\micro \meter}. Our grAl layer has a typical sheet resistance of $\unit{0.8}{\kilo \Omega}/\Box$ (see below for fabrication details), which corresponds to a sheet inductance of $\unit{0.7}{\nano\henry}/\Box$, assuming a critical temperature of \unit{1.85}{\kelvin}. With these parameters, simulations using the Sonnet software indicate that the impedance of the line is \unit{6}{\kilo \ohm} on a standard Silicon wafer. Note that this value may slightly differ from one sample to the other because of fluctuations in the grAl composition, which changes both the resistivity and the critical temperature.   

The DBR consists in two sections of microstrip lines with a large impedance mismatch. The first high impedance section is identical to the resonator, while the second section is made of Aluminum and has a much lower impedance of \unit{105}{\ohm}. The impedance mismatch between the two sections comes both, from the use of different materials, one with a large kinetic inductance and one without, and from the width difference, which is \unit{0.5}{\micro \meter} for the high impedance section and \unit{20}{\micro \meter} for the low impedance section. 

\begin{figure}[h!]
    \includegraphics[]{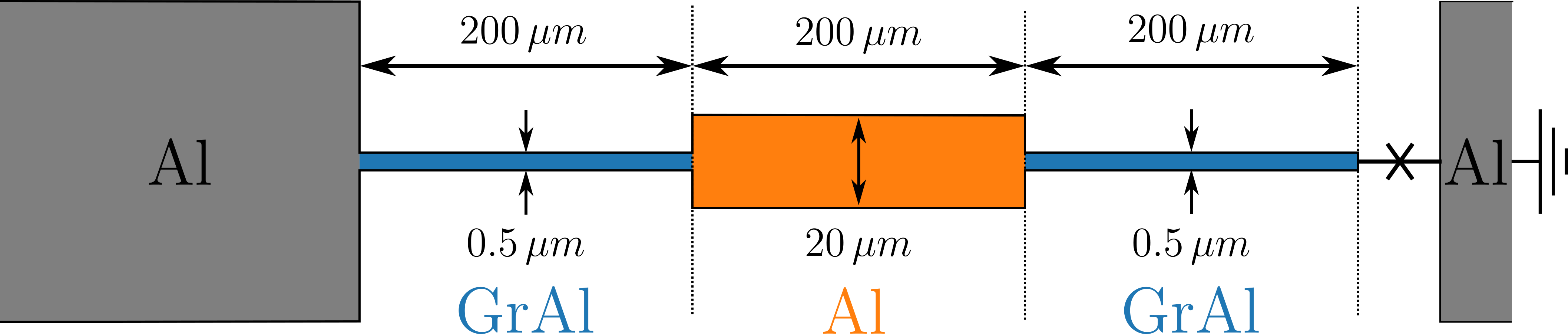}
    \caption{Geometry of the sample. The resonator and the first DBR section are made of GrAl and have a length that corresponds to one quarter wavelength at \unit{6}{\giga \hertz}. The second section is made of Al and has a much lower impedance. The junction connects the end of the resonator to the ground.}
    \label{fig:drawing_dbr}
\end{figure}

In the final design of the DBR, we reduce on purpose the length of the second (low impedance) section below the quarter wavelength Bragg condition. Otherwise, the quality factor of the fundamental mode of the resonator would be too high. The length is adjusted to target a quality factor of approximately \unit{1.3 \times 10^4}. Finally, an Al/AlOx/Al junction connects the left end of the resonator to the ground. A schematics of the sample is shown in figure \ref{fig:drawing_dbr}.

In order to evaluate the properties of the resonant modes of the system, we simulate the sample using the Sonnet software. The resonance frequency and the characteristic impedance of each mode are calculated from the admittance $Y(\omega)$ seen by the junction. We first numerically simulate the admittance matrix $Y_{ij}(\omega)$ of the two port network that is obtained by placing a first port at the position of the junction and a second one on the \unit{50}{\ohm} measurement line. From the admittance matrix, we then compute the admittance seen by the junction by eliminating the second port and taking into account the capacitance of the junction $C_J$. We arrive at
\begin{equation}
    Y(\omega) = Y_{11} - \frac{Y_{12} Y_{21}}{Y_{22} + Y_0} + i\omega C_J \, ,
\end{equation}
where $Y_0 = \unit{1/50}{\ohm^{-1}}$ is the admittance of the measurement line. 

%

The resonant frequencies $\omega_n$ of the modes are obtained by looking for the zeros of $\mathrm{Im}[Y(\omega)]$, while their characteristic impedance is given by \citep{Girvin2015}
\begin{equation}
    Z_{n} = \frac{2i}{\omega_n}\left( \frac{\partial Y(\omega)}{\partial \omega}\Bigr|_{\substack{\omega = \omega_n}} \right) ^{-1} \, .
    \label{eq:impedance_modes}
\end{equation}
In order to estimate $C_J$ and the precise value of the grAl sheet inductance, we compare the measured frequencies of the modes to the simulated ones. We find that a value of $C_J = \unit{1.75}{\femto \farad}$ and a sheet inductance of $\unit{0.56}{\nano\henry}/\Box$ explains well our results. It is compatible with the $~\unit{150 \times 150}{\nano \meter \squared}$ area of our junction. The measured values of the resonance frequencies are obtained from photo-assisted current measurements. The circuit is dc-biased below the gap and the current through the junction is measured as a function of the frequency of an incoming microwave signal. We observe sharp resonances that we identify to the modes coupled to the junction. 

In table \ref{tab:modes_adjusted_simulations_no_cap}, we report the measured resonance frequencies as well as the simulated properties of the modes. The mode with the largest impedance probed in the experiment corresponds to the fundamental mode of the quarter wavelength resonator, which is labeled with the index $n=1$. Because of the reduced length of the second section of the DBR with respect to the Bragg condition, there exists a lower frequency mode, which resonates at \unit{1.9}{\giga \hertz}. This mode has a relatively large impedance, even though it is not fully localized into the last high impedance section of the sample. The modes with even $n=2,4,6$ have a voltage minimum at the position of the junction and therefore a small impedance $Z_n$ as expected for a quarter wavelength resonator. After $n \geq 7$, the mode structure is affected by the DBR and becomes more complicated, we only list the modes that have a significant impedance. Note that the mode impedance as defined in (\ref{eq:impedance_modes}) is not an intrinsic property of the mode, as in the usual sense of a mode impedance. Here, it depends on the position of the junction and quantifies the vacuum quantum voltage fluctuation as seen by the junction. This definition coincides with the usual one when the junction is located at the maximum of voltage. 

Finally, we estimate the Kerr coefficient of the $n=1$ mode due to the non-linearity of the kinetic inductance. Test wires with the same geometry as the resonator and evaporated in the same conditions exhibit a critical current density of approximately \unit{1}{\milli \ampere . \micro \meter ^{-2}}. Following \citep{Maleeva2018}, we estimate from this measurement that the Kerr coefficient is approximately $2\pi \times \unit{200}{\hertz}$, which is negligible.

\begin{table}[h!]
    \centering
	\begin{tabular}{c c c c c}
		\hline
		& $\omega_{n, \mathrm{meas.}}/2\pi \, \unit{}{[\giga \hertz]}$ & $\omega_{n, \mathrm{sim.}}/2\pi \, \unit{}{[\giga \hertz]}$ & $Z_{n} \, \unit{}{[\kilo \ohm]}$ & $\lambda_{n}$  \\ 
		\hline
		n = 0 & 1.90 & 1.89 & 3.28 & 0.64 \\
		n = 1 & 5.91 & 5.91 & 4.48 & 0.74 \\
		n = 2 & 12.14 & 12.19 & 0.017 & 0.05 \\
		n = 3 & 15.83 & 15.54 & 1.66 & 0.45 \\
        n = 4 & 23.15 & 23.12 & \unit{2 \times 10^{-3}} & 0.01 \\
        n = 5 & & 25.34 & 1 & 0.35 \\
        n = 6 & & 34.06 &  \unit{<1 \times 10^{-3}} & $<0.01$ \\
		n = 7 & & 35.09 & 0.71 & 0.3 \\
		n = 8 & & 44.34 & 0.58 & 0.27 \\
		n = 10 & & 53.80 & 0.54 & 0.26 \\
		n = 12 & & 63.51 & 0.35 & 0.2 \\
		n = 14 & & 73.09 & 0.3 & 0.19 \\
		n = 16 & & 82.67 & 0.26 & 0.17 \\
		\hline
	\end{tabular}
	\caption{\label{tab:modes_adjusted_simulations_no_cap} Resonance frequency, characteristic impedance and displacement amplitude (see main text) of the resonator modes up to \unit{90}{\giga\hertz}. The second column corresponds to measured values, while the others are the results of numerical simulations. The values of $\omega$ and $Z_n$ for $n\geq 12$ are obtained by extrapolation.}
\end{table}

\section{Sample fabrication}
\noindent The sample is micro-fabricated on a \unit{300}{\micro \meter} thick, thermally oxidized, silicon substrate through the following steps.

\vspace{2mm}
\noindent\emph{Granular Aluminum resonator and DBR structure patterning}

\begin{figure*}
    \centering
    \begin{subfigure}[t]{0.32\textwidth}
        \centering
        \includegraphics[width=\textwidth]{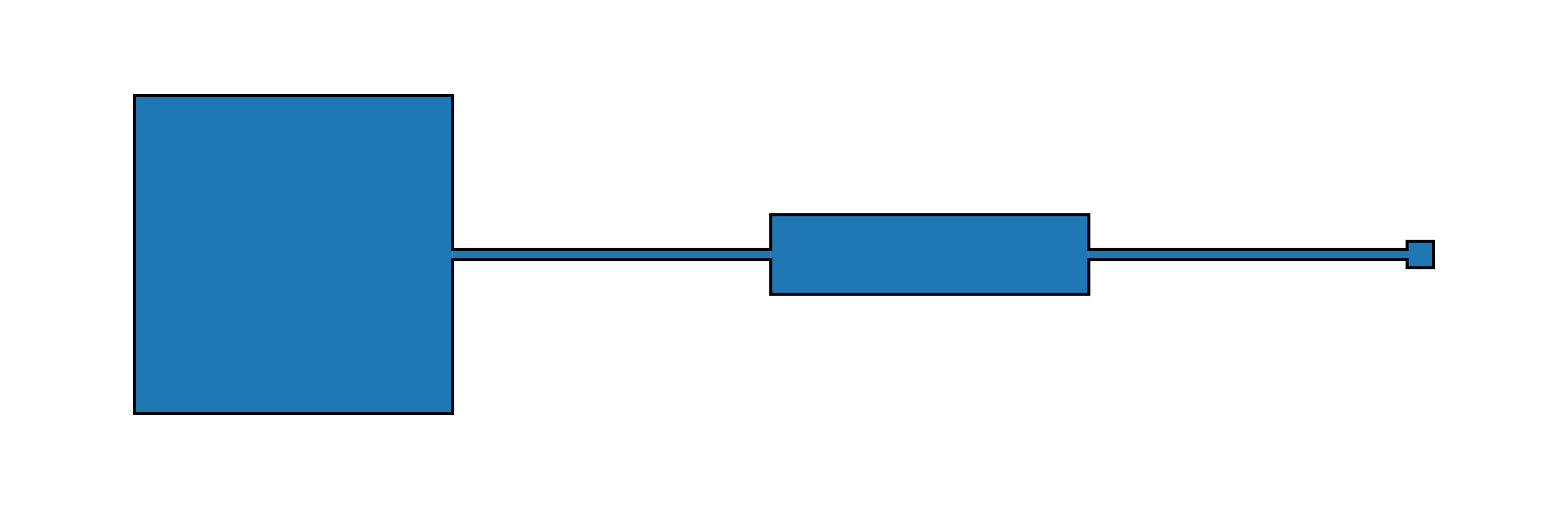}
        \caption{Evaporation of \unit{10}{\nano \meter} of grAl at \unit{0}{\degree} angle.}
        \label{fig:grAl_fabrication_flow_chart_1}
    \end{subfigure}
    \hfill
    \begin{subfigure}[t]{0.32\textwidth}  
        \centering 
        \includegraphics[width=\textwidth]{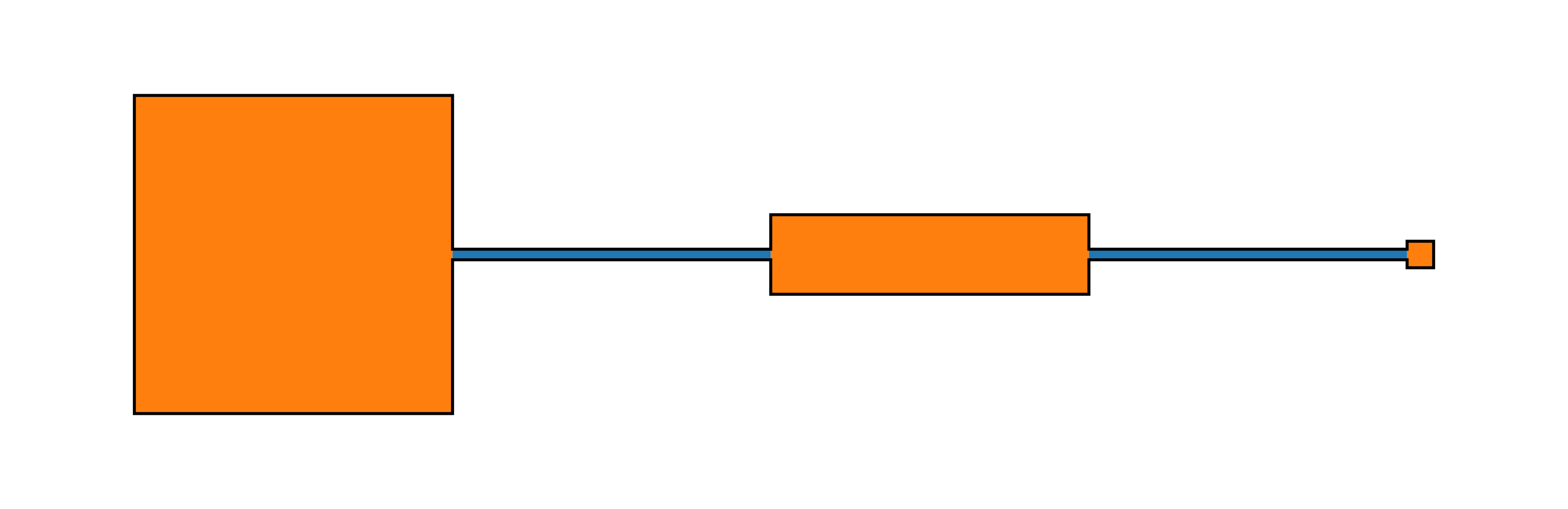}
        \caption{Evaporation of \unit{30}{\nano \meter} of Al at \unit{45}{\degree} angle.}
        \label{fig:grAl_fabrication_flow_chart_2}
    \end{subfigure}
    \begin{subfigure}[t]{0.32\textwidth}   
        \centering 
        \includegraphics[width=\textwidth]{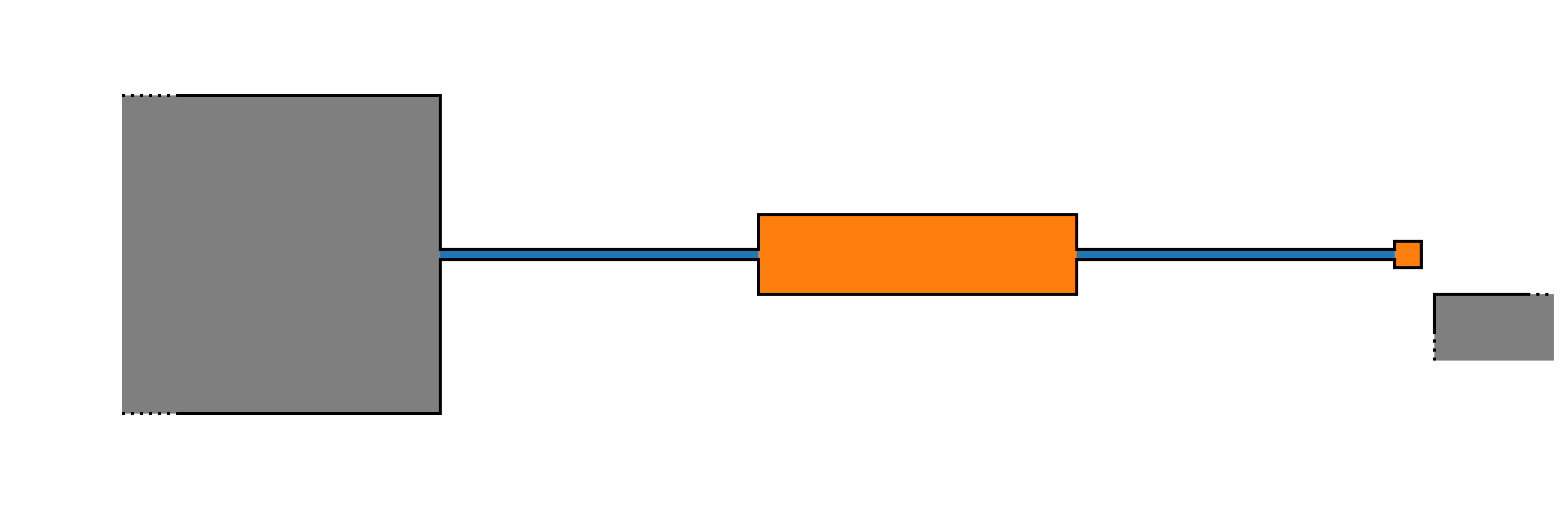}
        \caption{Patterning of measurement lines and ground contacts in Al.}
        \label{fig:grAl_fabrication_flow_chart_3}
    \end{subfigure}
    \vskip\baselineskip
    \begin{subfigure}[t]{0.32\textwidth}   
        \centering 
        \includegraphics[width=\textwidth]{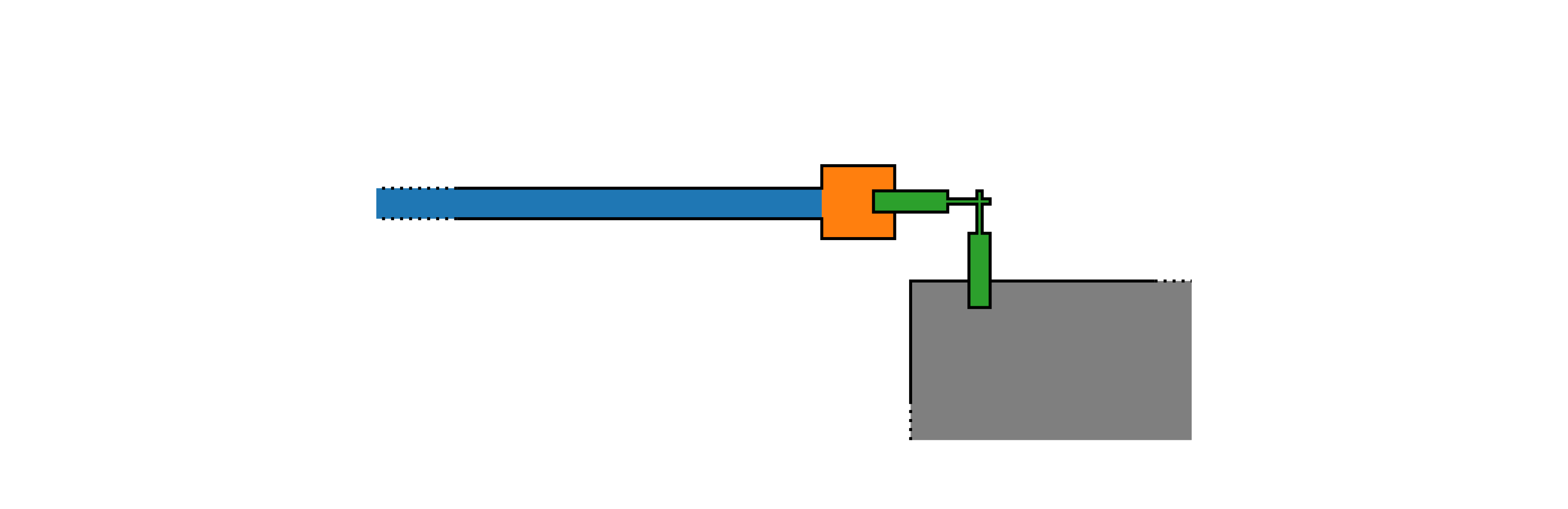}
        \caption{Realization of the Al/AlOx/Al tunnel junction.}
        \label{fig:grAl_fabrication_flow_chart_4}
    \end{subfigure}
    \begin{subfigure}[t]{0.32\textwidth}   
        \centering 
        \includegraphics[width=\textwidth]{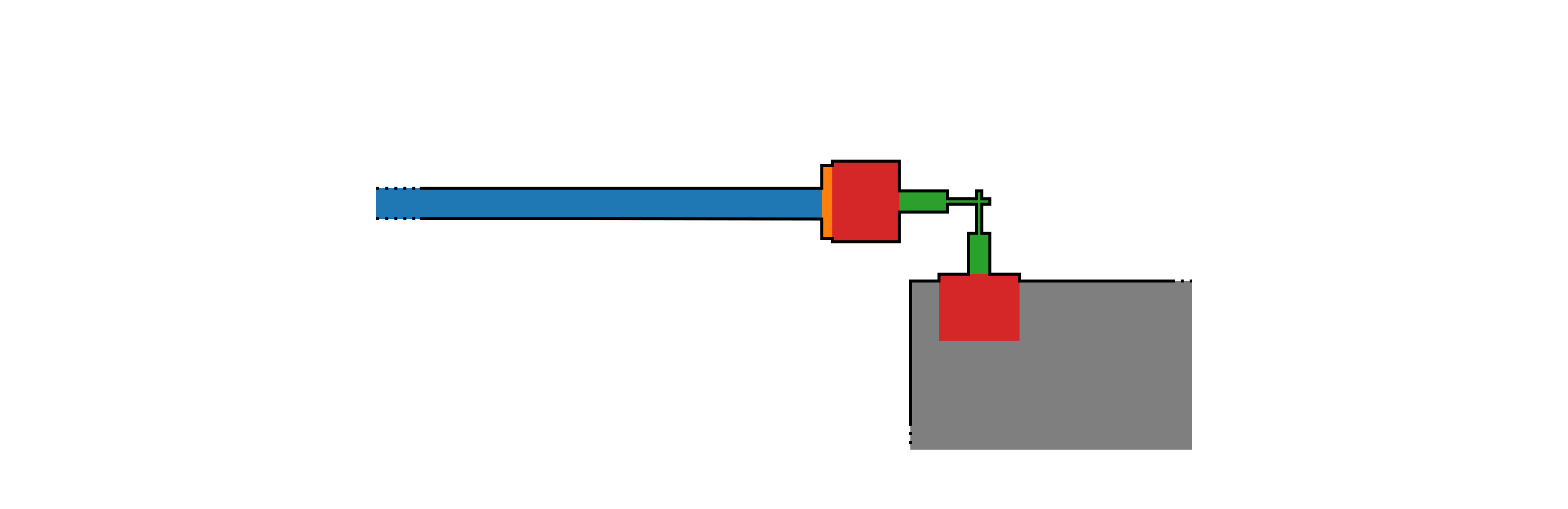}
        \caption{Connection of the junction contacts.}
        \label{fig:grAl_fabrication_flow_chart_5}
    \end{subfigure}
    \caption{Main steps of the fabrication process.} 
    \label{fig:grAl_fabrication_flow_chart}
\end{figure*}

The design is written on a \unit{1.3}{\micro \meter} thick trilayer resist stack (PMMA 495/PMMA 950/PMMA 950) using electronic beam lithography.
After revelation, \unit{10}{\nano \meter} of grAl are deposited by e-beam evaporation at \unit{0}{\degree} angle (fig. \ref{fig:grAl_fabrication_flow_chart_1}). The oxygen pressure during the evaporation is adjusted to \unit{1.1 \times 10^{-5}}{\milli \bbar} and the evaporation rate is \unit{1.5}{\angstrom \per \second}. In a following evaporation, without exposing the sample to air, we deposit \unit{30}{\nano \meter} of pure Al with an angle of \unit{45}{\degree} in the direction perpendicular to the one of the wires (fig. \ref{fig:grAl_fabrication_flow_chart_2}). Because of the the resist thickness and the evaporation angle, only regions with a width exceeding \unit{1.3}{\micro \meter} are covered during this second evaporation. This corresponds, from left to right, to the measurement line, the second DBR section and a final \unit{2.5x2.5}{\micro \meter\squared} square contact that is used to connect the junction. 

In a second step, we pattern the remaining part of the \unit{50}{\ohm} measurement line and a central ground plane, which is used to connect the other side of the junction (fig. \ref{fig:grAl_fabrication_flow_chart_3}). These structures are realized through optical lithography and liftoff of a \unit{60}{\nano \meter} thick Al layer. In order to obtain a good dc contact with the previously deposited measurement line, an argon ion beam etching step is used before the evaporation. At the end of the process a \unit{30}{\nano \meter} layer of Nb is deposited at the back of the sample for the ground plane.

\vspace{2mm}
\noindent\emph{Tunnel junction patterning}

The Al/AlOx/Al junction is realized through a standard double angle evaporation using a PMMA bilayer resist without any suspended bridge (''Manhattan technique``, see fig. \ref{fig:grAl_fabrication_flow_chart_4}). The width of the junction arms is \unit{150}{\nano \meter} and the oxidation is performed at \unit{10}{\milli \bbar} oxygen pressure (static) for 20\,minutes.
In a final step, we connect the junction contacts to both the resonator and the ground plane using a last step of optical lithography, followed by Ar ion beam etching and the evaporation of \unit{80}{\nano \meter} of Al (fig. \ref{fig:grAl_fabrication_flow_chart_5}). 

\section{Experimental setup}
The sample is glued on a gold-plated copper sample holder and bonded to a microwave PCB.
The sample holder is fixed on the \unit{10}{\milli \kelvin} stage of a Cryoconcept dry dilution refrigerator. 
\begin{figure}[ht!]
    \includegraphics[width = 0.45\linewidth]{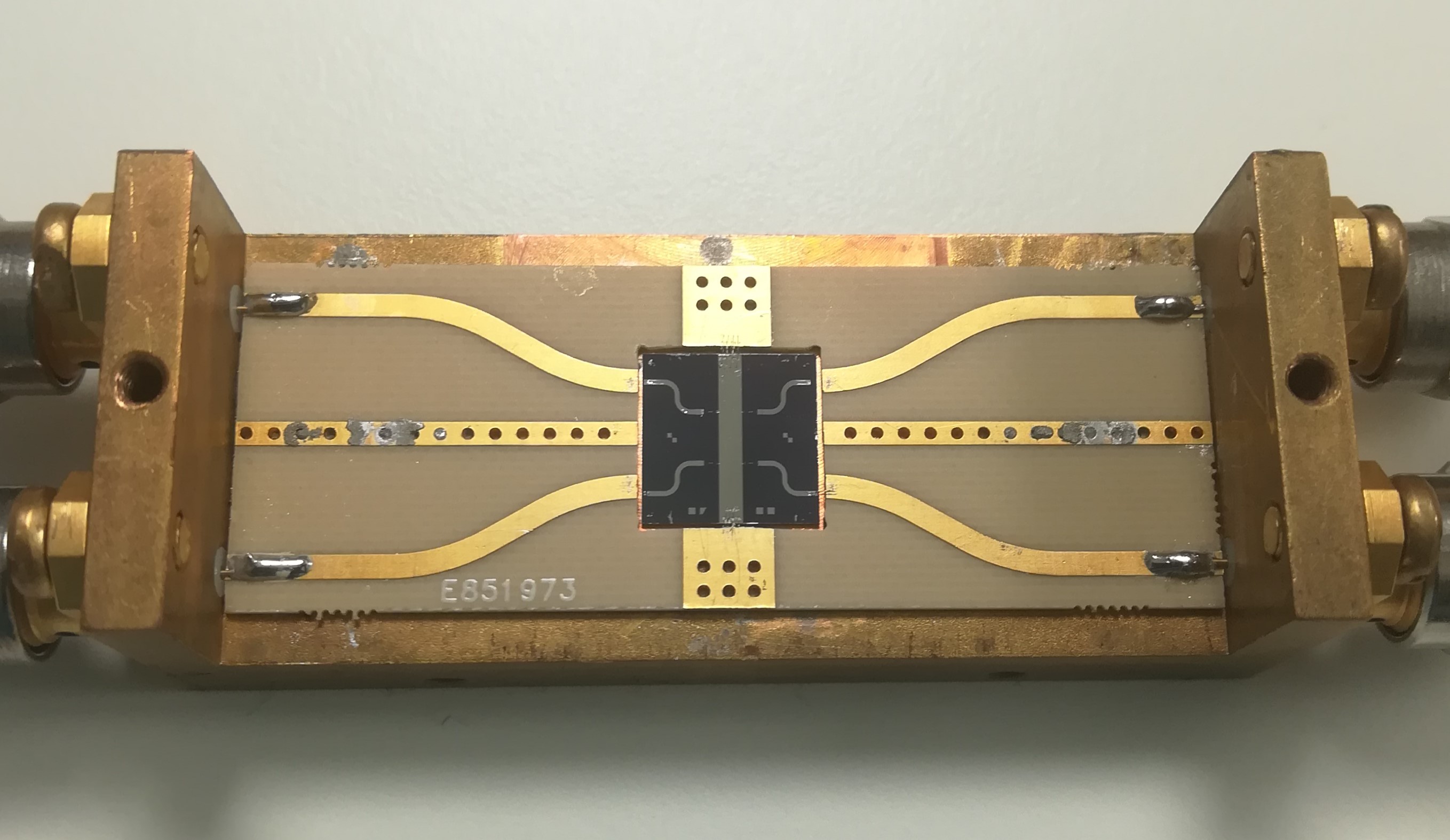}
    \caption{Picture of the sample holder and of the PCB to which the sample is wire bonded.}
    \label{fig:sample_holder}
\end{figure}

\begin{figure}[ht!]
    \includegraphics[]{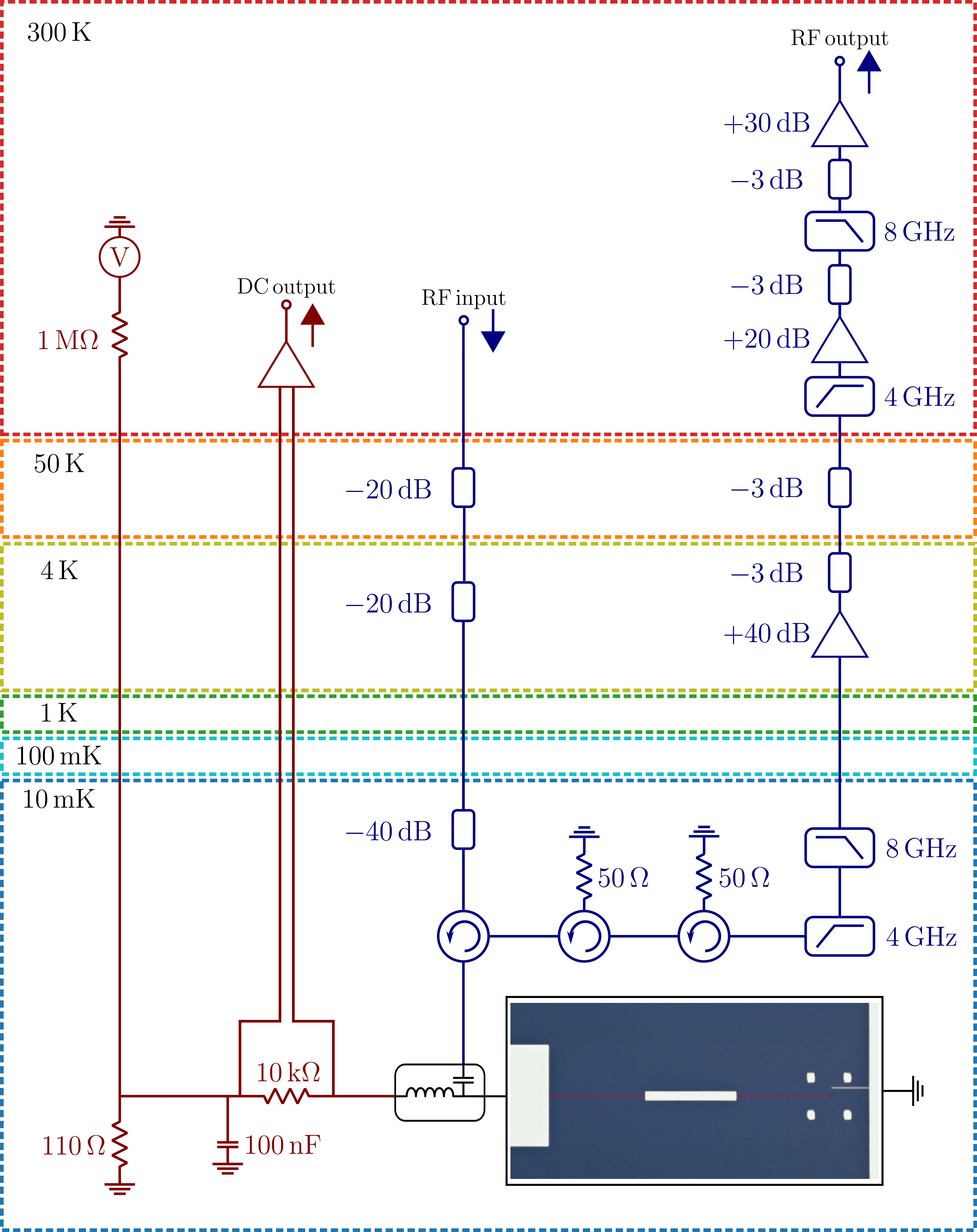}
    \caption{Microwave and dc setup used for the experiment.}
    \label{fig:experimental_setup}
\end{figure}

The experimental setup used to measure the sample is shown in figure \ref{fig:experimental_setup}. All the microwave measurements presented in the manuscript are performed with a vector network analyzer connected to the RF input and output ports. The microwave drive is sent to the sample through an attenuated coaxial line and its reflection is collected by a different line using a circulator. The reflected signal is amplified by a cryogenic HEMT amplifier at \unit{4}{\kelvin} followed by two room temperature amplifiers. 
In order to dc-bias the junction, we use a voltage divider followed by a filtering capacitor and a bias tee. A \unit{10}{\kilo \ohm} resistance is mounted in series with the sample to measure the current through the junction.


\section{Junction admittance}
\begin{figure}[ht!]
    \includegraphics[]{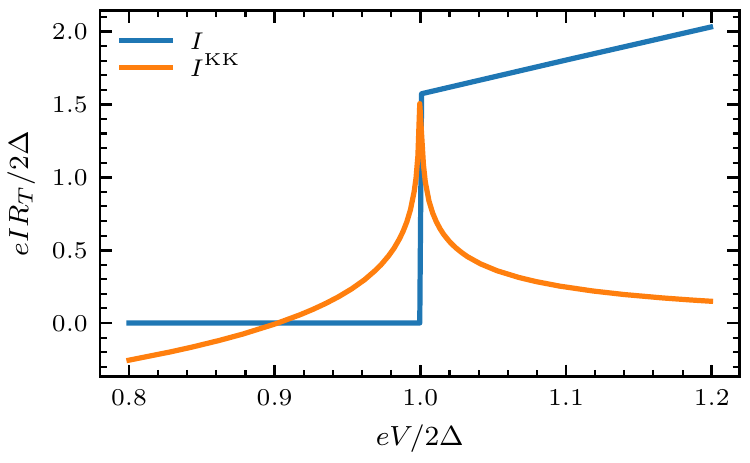}
    \caption{Quasiparticle current  and its Kramers-Kronig transform near the superconducting gap.}
    \label{fig:QP_current}
\end{figure}

When the impedance of the mode coupled to the junction is small compared to the quantum of resistance ($\lambda \ll 1$), the resonator field can be treated classically. In this case, the effect of charge tunneling can be modeled by treating the junction as a frequency and voltage dependent admittance $Y_J$ \cite{tucker_quantum_1985}. This admittance also varies with the microwave amplitude seen by the junction. Supposing a weak microwave amplitude, the real and imaginary parts of the admittance can be expanded to first order in the microwave amplitude as \citep{tucker_quantum_1985,worsham_quantum_1991}:  
\begin{align}
    \mathrm{Re}(Y_J) & = \frac{e}{2\hbar \omega} \left[ I(V + \hbar\omega/e) -  I(V - \hbar\omega/e) \right] 
    \label{eq:admittance_junction_real} \\
    \mathrm{Im}(Y_J) & = \frac{e}{2\hbar \omega} \left[ I^{\KK}(V + \hbar\omega/e) + I^{\KK}(V - \hbar\omega/e) -2I^{\KK}(V)\right] \, .
    \label{eq:admittance_junction_imag}
\end{align}
Here, $I$ is the quasiparticle current that is defined as
\begin{equation}
    I(V) = \frac{1}{R_T}\int_{-\infty}^\infty n_L(V')n_R(V'-V)[f(V'-V)-f(V')] \, dV' \, ,
\end{equation}
where $n_R$ ($n_L)$ is the dimensionless density of states in the right (left) contact, which tend to one at voltages far below and far above the gap, and $f(V)$ is the Fermi distribution. 
Finally, $I^{\KK}(V)$ is the Kramers-Kronig transform of $I(V)$ as defined in the main text. Their variations near the superconducting gap $V=2\Delta/e$ are shown in figure \ref{fig:QP_current}.

Taking into account this admittance in the calculation of the resonator mode properties, the real part of the admittance translates into a loss rate $\kappa$ and the imaginary part into a resonance frequency shift $\delta \omega$. The two quantities are given by
\begin{align}
    \kappa & =  \frac{\lambda^2}{e} \left[ I(V + \hbar\omega/e) -  I(V - \hbar\omega/e) \right] \\
    \delta \omega & = -\frac{\lambda^2}{2e} \left[ I^\KK(V + \hbar\omega/e) + I^\KK(V - \hbar\omega/e) -2I^\KK(V)\right] \, . 
\end{align}
These expressions can be derived using equations (2) and (3) of the main text in the limit of small $\lambda$. In a quantum approach, they correspond to the usual single photon loss and frequency shift that arise when an harmonic oscillator mode is linearly coupled to a bath. 
In the strong impedance regime, where $\lambda \approx 1$, this approach is not valid, and a quantum treatment of the field, and in particular of the displacement operator appearing in the tunnel Hamiltonian, is necessary.

\section{Quantum master equation}
\label{sec:master_equation}
In order to describe the quantum dynamics of the resonator field in the high impedance regime, we use a master equation formalism and treat the junction as a bath and the resonator as an open quantum system. Following the standard quantum optics approach based on the Born-Markov and on the secular approximation \citep{breuer_theory_2007}, one obtains a master equation in the Lindblad form for the density matrix of the resonator modes as detailed, for example, in \citep{esteve_quantum_2018}. 
The master equation is written in terms of jump operators $\hat{A}_{l}$ operators  that come from the expansion of the displacement operator $\mathrm{exp}[i\lambda(\hat{a} + \hat{a}^\dagger)]$ in the Fock state basis. Their effect is to create $l$ photons when $l>0$ and annihilate $-l$ photons when $l<0$. They are defined, for positives $l$, as
\begin{equation}
    \hat{A}_{l>0} = \sum_{n=0}^\infty \ket{n+l} \bra{n+l} e^{i \lambda (\hat{a}+\hat{a}^\dagger)} \ket{n} \bra{n} \, \label{eq.Al}.
\end{equation}
The operators with $l<0$ are defined through $\hat{A}_{-l} = (-1)^{l} \hat{A}_{l}^{\dagger}$. For $l\geq 0$, the matrix elements of $\hat{A}_{l}$ can be obtained from the following identity
\begin{equation}
    \mel{n+l}{e^{i \lambda (\hat{a}+\hat{a}^\dagger)}}{n} = \sqrt{\frac{n!}{(n+l)!}} e^{-\lambda^2/2} (i\lambda)^l L_n^{(l)}(\lambda^2)\,, \label{eq.wnl}
\end{equation}
where $L_n^{(l)}$ are the generalized Laguerre polynomials. In the main text, we define $\alpha_{nl} = |\mel{n+l}{e^{i \lambda (\hat{a}+\hat{a}^\dagger)}}{n}|^2$, which can be rewritten in terms of $\hat{A}_{l}$ operators as $\alpha_{nl}=|\mel{n+l}{A_l}{n}|^2$ for $l \geq 0$.
When the non-linearity is strong ($\lambda \approx 1$), quantum jumps, generated by the $\hat{A}^\dagger_{l}$ operators, lead to transitions from $\ket{n}$ to $\ket{n-l}$. 
This is in stark contrast to the usual quantum optics situation, where the two possible jump operators are $\hat{a}$ and $\hat{a}^\dagger$, which only couples adjacent Fock states.
In the case where many modes are coupled to the junction, jump operators can be defined for each mode using the value of $\lambda$ obtained from numerical simulations of the mode impedance seen by the junction (see \ref{sec:emsimulation}). 
In order to obtain more compact expressions, we define
\begin{equation}
    \hat{A}_{l_0,l_1, \ldots } = \hat{A}_{l_0} \otimes \hat{A}_{l_1} \otimes \ldots \, ,
\end{equation}
where $\hat{A}_{l_n}$ is the operator defined in (\ref{eq.Al}) associated to mode $n$. 
The other ingredients entering the master equation are the real and imaginary part of the single sided Fourier transform of the bath correlation function. In the case of a tunnel junction, one obtains 
\begin{align}
    \gamma(V) & =  \frac{1}{eR_T}\int_{-\infty}^\infty n_R(V')n_L(V+V')f(V')(1-f(V+V')) \, dV' \\
    \epsilon(V) & = \frac{1}{2\pi} \mathcal{P} \int_{-\infty}^\infty \frac{\gamma(V')}{V-V'} \, dV' \, .
\end{align}
With these definitions, the master equation for the density matrix of the modes is
\begin{align}
    \frac{d\rho}{d t} & = -i \comm{ \hat{H}_\mathrm{LS}}{\rho} + \sum_{l_0,l_1,\ldots = -\infty}^{\infty}  \left[ \gamma(V + \sum_n l_n \hbar \omega_n/e) + \gamma(-V + \sum_n l_n \hbar \omega_n/e) \right] \mathcal{D}[\hat{A}^\dagger_{l_0,l_1, \ldots }](\rho) \label{eq:master_equation}\\
    \hat{H}_\mathrm{LS} & = \sum_{l_0,l_1,\ldots = -\infty}^{\infty}  \left[ \epsilon(V + \sum_n l_n \hbar \omega_n /e) + \epsilon(-V + \sum_n l_n \hbar \omega_n/e) \right] \hat{A}_{l_0,l_1, \ldots } \hat{A}^\dagger_{l_0,l_1, \ldots } \\
     \mathcal{D}[\hat{X}](\rho) & = \hat{X}\rho \hat{X}^\dagger - \frac{1}{2}\acomm{\hat{X}^\dagger\hat{X}} {\rho} \, .
\end{align}
The first term of equation (\ref{eq:master_equation}) describes the Lamb shift of the resonator energy levels with associated Hamiltonian $H_\mathrm{LS}$, while the second term corresponds to photo-assisted tunneling events, during which one electron tunnels and photons are absorbed and/or created in the resonator modes. As required by gauge invariance, the master equation is unchanged by $V \rightarrow -V$. While deriving this equation, we assume that the mode frequencies are not commensurate when performing the secular approximation. The bath functions $\gamma(V)$ and $\epsilon(V)$ can be related to the $I(V)$ characteristic of the junction by noticing that
\begin{alignat}{2}
    I(V) & = e [ \gamma(V) - \gamma(-V)]  & & \approx  e \, \gamma(V) \\
    I^\KK(V) & =  -2e [ \epsilon(V) + \epsilon(-V)] & & \approx -2e\, \epsilon(V) \,.
\end{alignat}
The last approximations are valid for positive bias and low temperature. With these approximations, the master equation can be rewritten as
\begin{align}
    \frac{d\rho}{d t} & = -i \comm{\hat{H}_\mathrm{LS}}{\rho} + \frac{1}{e} \sum_{l_0,l_1,\ldots = -\infty}^{\infty} I(V + \sum_n l_n \hbar \omega_n/e) \mathcal{D}[\hat{A}^\dagger_{l_0,l_1, \ldots }](\rho) \\
    \hat{H}_\mathrm{LS} & = -\frac{1}{2e} \sum_{l_0,l_1,\ldots = -\infty}^{\infty}  I^\KK(V + \sum_n l_n \hbar \omega_n /e) \hat{A}_{l_0,l_1, \ldots } \hat{A}^\dagger_{l_0,l_1, \ldots } \, .
\end{align}
This master equation is only valid for positive bias and as long as the energy shift $\sum_n l_n \hbar \omega_n$ remains small compared to the gap $2 \Delta$, which is the case in our experiment. 
This multimode master equation is in general difficult to solve. In the quantum Zeno experiment presented in the main text, only the \unit{6}{\giga\hertz} mode is pumped. We therefore assume that the other modes do not participate to the dynamics and remain in vacuum. When tracing out these modes, we have to compute
\begin{align}
    \Tr_{0,2,\ldots} \left[ \hat{A}_{l_0,l_1, \ldots} \hat{A}^\dagger_{l_0,l_1, \ldots} \rho \right] & = \expval{\hat{A}_{l_0} \hat{A}^\dagger_{l_0}}{0} \expval{\hat{A}_{l_2} \hat{A}^\dagger_{l_2}}{0}  \ldots  \hat{A}_{l_1} \hat{A}^\dagger_{l_1} \rho_1 \\
    & = |\mel{0}{\hat{A}_{l_0}}{-l_0}|^2 \, |\mel{0}{\hat{A}_{l_2}}{-l_2}|^2\ldots  \hat{A}_{l_1} \hat{A}^\dagger_{l_1} \rho_1 \\
    & = \begin{cases}
    W_{l_0} e^{-\lambda_0^2}  \, W_{l_2} e^{-\lambda_2^2}   \ldots \hat{A}_{l_1} \hat{A}^\dagger_{l_1} \rho_1 \ \ \mathrm{if} \ l_0 \leq 0, \, l_2\leq 0, \ldots \\
    0 \ \ \mathrm{otherwise} \, ,
    \end{cases}
\end{align}
with $W_{l_n} = \lambda_n^{-2 l_n}/(-l_n)! \,$. The master equation for the $n=1$ mode is then given by
\begin{align}
    \frac{d\rho_1}{d t} & = -i \comm{\hat{H}_\mathrm{LS}}{\rho_1} + \frac{1}{e} \sum_{l_1 = -\infty}^{\infty} \ \sum_{l_0,l_2, \ldots = -\infty}^{0}  W_{l_0} W_{l_2} \ldots \tilde{I}(V + \sum_n l_n \hbar \omega_n/e) \mathcal{D}[\hat{A}^\dagger_{l_1}](\rho_1) \label{eq:master_Wl} \\
     \hat{H}_\mathrm{LS}  & = -\frac{1}{2e} \sum_{l_1 = -\infty}^{\infty} \ \sum_{l_0,l_2, \ldots = -\infty}^{0}  W_{l_0} W_{l_2} \ldots \tilde{I}^\KK( V + \sum_n l_n \hbar \omega_n /e )  \hat{A}_{l_1} \hat{A}^\dagger_{l_1}  \, ,
\end{align}
The modified current characteristic $\tilde{I}$ takes into account the dynamical Coulomb blockade (DCB) due to the other modes. It is defined as
\begin{equation}
    \tilde{I}(V) = e^{-\lambda_0^2} e^{-\lambda_2^2} e^{-\lambda_3^2} \ldots   I(V) \, .
\end{equation}
The multi-mode DCB factor is the product of the blockade factors $|\expval{\exp[i\lambda_n(\hat{a}_n+\hat{a}^\dagger_n)]}{0}|^2 = e^{-\lambda_n^2}$ for each mode, except the \unit{6}{\giga\hertz} mode.
As mentioned in the main text, this is equivalent to an increase of the tunnel resistance by a factor $\Pi_{n \neq 1} e^{\lambda_n^2}$.

\begin{figure}[h!]
    \centering
    \includegraphics{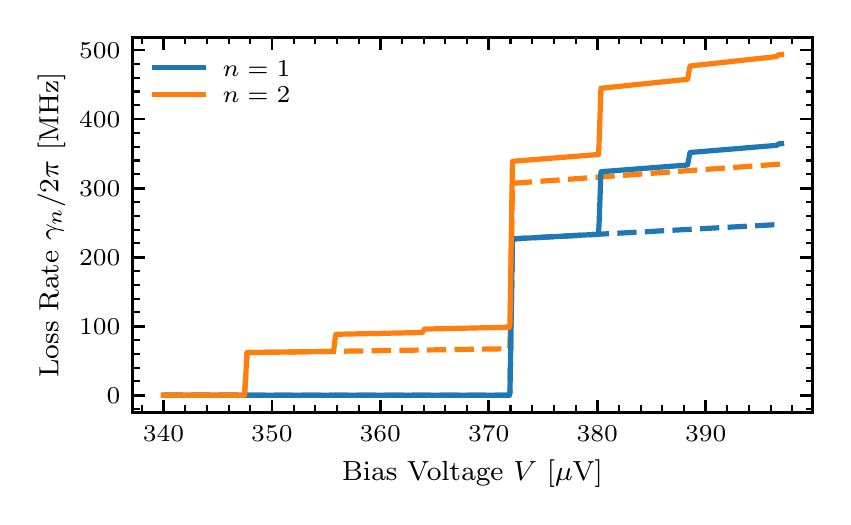}
    \caption{Comparison of the different predictions for the loss rates of the first two Fock states $n=1$ and $n=2$ as a function of voltage. The dashed lines correspond to  predictions of (\ref{eq:master_effective}), as plotted in figure 2a of the main text. Solid lines correspond to the predictions of (\ref{eq:master_Wl}).}
    \label{fig:Gamma_n_approx}
\end{figure}

The master equation can be further simplified by keeping only the $l_n=0$ term in the sum over the other modes. This assumes that the effect of these modes is simply to renormalize the tunnel resistance. We then obtain
\begin{align}
    \frac{d\rho_1}{d t} & = -i \comm{\hat{H}_\mathrm{LS}}{\rho_1} + \frac{1}{e} \sum_{l_1 = -\infty}^{\infty} \tilde{I}(V +  l_1 \hbar \omega_1/e)  \mathcal{D}[\hat{A}^\dagger_{l_1}](\rho_1) \label{eq:master_effective}\\
     \hat{H}_\mathrm{LS}  & = -\frac{1}{2e} \sum_{l_1 = -\infty}^{\infty} \tilde{I}^\KK( V + l_1 \hbar \omega_1 /e )  \hat{A}_{l_1} \hat{A}^\dagger_{l_1}   \, ,
\end{align}
For voltages below the gap, the simplification of the jump terms only neglects the contribution of the \unit{1.9}{\giga\hertz} mode. In particular, we neglect processes where photons in the \unit{6}{\giga\hertz} are converted into photons at \unit{1.9}{\giga\hertz} and the remaining energy is taken by the tunneling electron (see figure \ref{fig:Gamma_n_approx}). The higher frequency modes have no influence because of energy conservation. Computing the loss rate for state $\ket{n}$ using (\ref{eq:master_effective}) leads to equation (2) of the main text. The expression of $\hat{H}_\mathrm{LS}$ is more affected than the jump terms, because all the modes contribute as discussed in more details below (see \ref{sec:LambShift}).



\section{Zeno dynamics and pump calibration}
We now use the previously derived master equation to explain the quantum Zeno dynamics observed in the experiment.
Quantum Zeno dynamics happens when we dc-bias the junction at \unit{363}{\micro \volt}, which lies in the voltage region where the $l \geq 2$ processes are the dominant loss process. We measure the resonator reflected spectrum $S_{11}$ for different microwave powers (see inset of figure 3b of the main text) and extract the mode amplitude from these measurements.
This requires to know the pump amplitude, which is defined as $\eta=\sqrt{\kappa_c P/\hbar \omega}$, where $\kappa_c$ is the coupling loss rate and $P$ is the incoming microwave power on the sample. The coupling $\kappa_c$ is measured independently. The incoming power $P$ depends on the total attenuation of the experimental setup that needs to be calibrated. We do so by measuring the resonance fwhm and by comparing it to the one of a two level system. 
In a two level system, the fwhm is equal to $\sqrt{\kappa^2 + 4\Gamma^2}$, with $\kappa$ being the total loss rate and $\Gamma$ the power broadening, which varies with the pump amplitude $\eta$ as $\Gamma = \sqrt{2}\eta$.
By assuming the same power dependence on the measured $\Gamma$, we arrive at the attenuation of the line that pumps the resonator. 

\begin{figure}[ht!]
    \centering
    \includegraphics{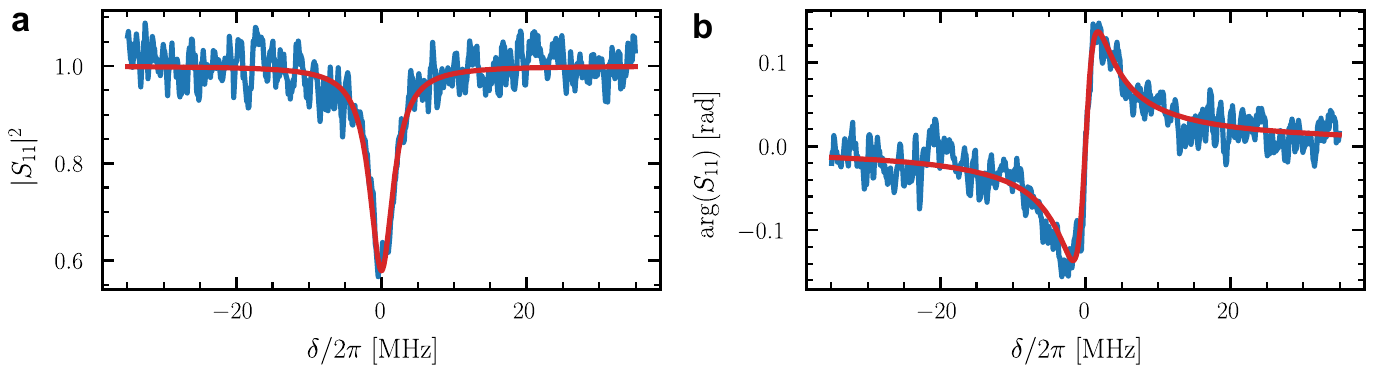}
    \caption{ {\bf a} Squared modulus and {\bf b} phase of the reflection spectrum measured at $P=\unit{-162}{dBm}$. The red curve shows the fitted spectrum assuming a two-level system model.}
    \label{fig:fit_spectrum}
\end{figure}

\begin{figure}[ht!]
    \centering
    \includegraphics{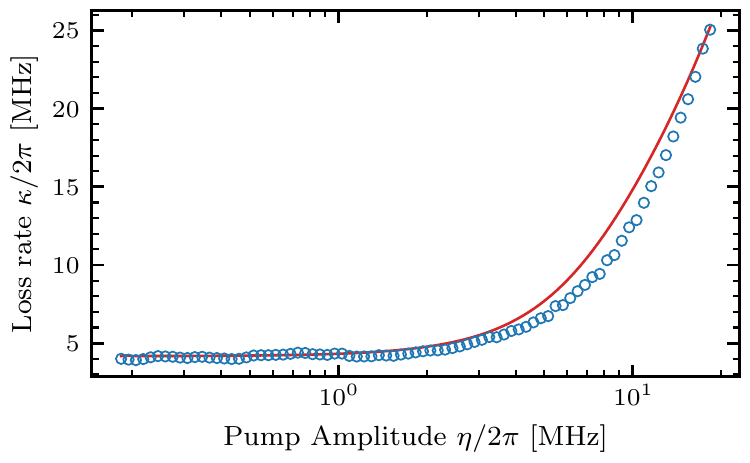}
    \caption{Comparison between the measured (blue data) and simulated (red curve) evolution of the loss rate $\kappa$. The curves show that $\kappa$ remains constant, as in a two level system, as long as the Zeno effect limits the dynamics of the system to the first two Fock states. When the Zeno dynamics break down, the loss rate increases quadratically with the pump amplitude.}
    \label{fig:LossRate_ME}
\end{figure}

We extract the loss rate $\kappa$ and the broadening $\Gamma$ by fitting the spectra measured at different power using the two-level formula \citep{CohenTannoudji1998}
\begin{equation}
    S_{11} = 1 - \kappa_c\frac{\kappa/2 - i\delta}{(\kappa/2)^2 + \delta^2 + \Gamma^2} \, ,
\end{equation}
where $\delta$ is the frequency detuning between the pump and the $0\rightarrow 1$ resonance. In figure \ref{fig:fit_spectrum}, we plot the spectrum measured at $P=\unit{-162}{dBm}$ together with the fitted $S_{11}$ (red curve). This procedure is repeated for every power, and by fitting the slope of $\Gamma$ with power, we obtain the attenuation and the value of $\eta$ as used in figure 3 of the main text. 
Once the pump is calibrated, we extract the resonator field intensity from the measured spectra through
\begin{equation}
   \langle \hat{a} \rangle  = \frac{\eta}{\kappa_c} \left( 1 -  S_{11} \right) \, .    
\end{equation}
A plot of $|\langle \hat{a} \rangle|^2$ at resonance is shown in figure 3a of the main text. We compare this measured intensity to the predictions of the following master equation
\begin{align}
    \frac{d\rho}{d t} &  = -i \comm{\hat{H}}{\rho} + \kappa \mathcal{D}[\hat{a}](\rho) + \frac{1}{e} \sum_{l= -\infty}^{\infty} \tilde{I}(V + l \hbar \omega/e) \mathcal{D}[\hat{A}^\dagger_{l}](\rho) \label{eq:mastereq}\\
    \hat{H} & = i\eta(\hat{a} - \hat{a}^\dagger) -\delta \, \hat{a}^\dagger \hat{a} + \hat{H}_\mathrm{LS} \label{eq:Hamiltonian}\\
    \hat{H}_\mathrm{LS} & = -\frac{1}{2e} \sum_{l= -\infty}^{\infty}  \tilde{I}^\KK(V + l \hbar \omega /e) \hat{A}_{l} \hat{A}^\dagger_{l} \, .
\end{align}
This master equation is the simplified single mode master equation obtained in (\ref{eq:master_effective}), where we add a pump term with amplitude $\eta$, a detuning term proportional to $\delta$, and also include a single photon loss rate $\kappa$ in order to take into account single photon loss from various origins. This loss rate is $\kappa = 2\pi \times \unit{4.2}{\mega\hertz}$ ($\kappa = 2\pi \times \unit{2.7}{\mega\hertz}$) for $V=\unit{363}{\micro\volt}$ ($V=\unit{284}{\micro\volt}$). These values are chosen to reproduce the observed resonance width at low pump power. The expectation value of $\hat{a}$ is then obtained from the steady state solution of the master equation (calculated using the QuTip Python software \citep{Johansson2013}) and allows us to plot the curve in figure 3a of the main text. We also compute the expected $S_{11}$ and apply the same fitting procedure as for the experimental data and obtain the theoretical power broadening $\Gamma$ (shown in figure 3b of the main text) and the loss rate $\kappa$ (plotted in figure \ref{fig:LossRate_ME}). 

\begin{figure}[ht!]
    \centering
    \includegraphics{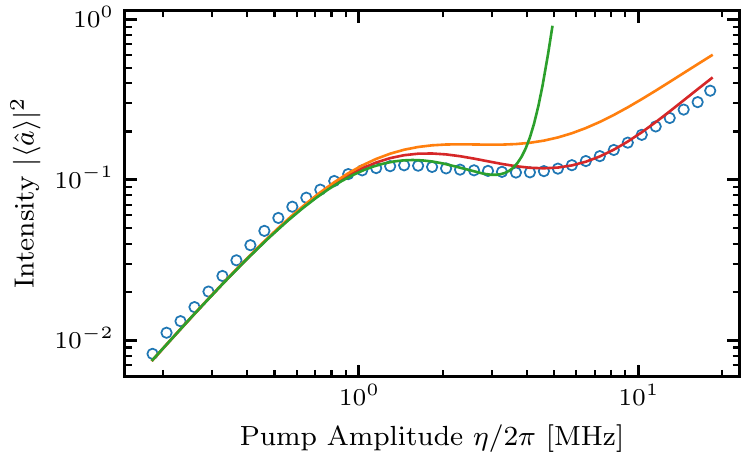}
    \caption{Comparison of the mode intensity predicted by different master equations. The red (orange) curve corresponds to the solution of (\ref{eq:mastereq}) with (without) the Lamb shift term at $V=\unit{363}{\micro\volt}$. The green curve shows the prediction for an ideal projective measurement in the $\{ \ket{0}, \ket{1}\}$ subspace. The blue points correspond to the experimental results shown in figure 3 of the main text.}
    \label{fig:LSnoLS}
\end{figure}

In figure \ref{fig:LSnoLS}, we compare the results of simulations including or not the Lamb shift term $H_\mathrm{LS}$. The non-linearity introduced by the Lamb shift significantly contributes to the blockade of the $1\rightarrow 2$ transition in addition to the Zeno effect. The figure of merit of the observed blockade can be quantified by the range of pump amplitude over which the mode behaves as a saturated two level system. Supposing that the mode is subject only to one and two photon loss, without any non-linearity, this range is set by the ratio $\tilde{\gamma}_2/\tilde{\gamma}_1$, where $\tilde{\gamma}_n$ is the effective loss rate of state $\ket{n}$ in this model. We find that, without Lamb shift, the ratio $\tilde{\gamma}_2/\tilde{\gamma}_1$ is about 15, with $\tilde{\gamma}_2$ being $2\pi \times \unit{65}{\mega\hertz}$ and $\tilde{\gamma}_1 = 2\pi \times \unit{4.2}{\mega\hertz}$. When we include the Lamb shift $\tilde{\gamma}_2$ increases approximately to $2\pi \times \unit{100}{\mega\hertz}$ with the ratio $\tilde{\gamma}_2/\tilde{\gamma}_1$ being about 25.

Finally, we simulate a master equation modeling ideal Zeno dynamics (see figure \ref{fig:LSnoLS}). We suppose that the Hamiltonian is given by equation (\ref{eq:Hamiltonian}) without the Lamb shift term. We then include two jump operators. The first one describes single photon loss with rate $\kappa$ and the second one is the projector $\ketbra{0} + \ketbra{1}$, which measures if the state is in the $\{ \ket{0}, \ket{1}\}$ subspace. The projection rate $\gamma_P$ is adjusted to fit the data in the region $\eta < 2\pi \times \unit{3}{\mega \hertz}$, we obtain $\gamma_P=2\pi \times \unit{215}{\mega \hertz}$, which is comparable to the rate of $2\pi \times \unit{100}{\mega \hertz}$ found above.

\section{Lamb Shift}
\label{sec:LambShift}
From the expression of $\hat{H}_\mathrm{LS}$ derived in \ref{sec:master_equation}, the Lamb shift $\delta\omega_n$ of the $\ket{n}$ state of the \unit{6}{\giga\hertz} mode, with all the other modes in vacuum, is given by
\begin{equation}
    \delta \omega_n = -\frac{1}{2e}\sum_{l_1 = -\infty}^{\infty} \ \sum_{l_0,l_2, \ldots = -\infty}^{0} W_{l_0} W_{l_2} \ldots \tilde{I}^\KK( V + \sum_n l_n \hbar \omega_n /e )   \matrixelement{n}{\hat{A}_{l_1} \hat{A}^\dagger_{l_1}}{n} \label{eq:LS1}\,.
\end{equation}
As discussed above, the main effect of the other modes is to renormalize the tunnel resistance, but the remaining terms modify the $V$ dependence beyond a simple renormalization of $R_T$. 
Including all the modes presented in \ref{sec:emsimulation}, we use this expression to compute the shift of the $\ket{0}$, $\ket{1}$ and $\ket{2}$ states, from which we deduce the curves plotted in figure 4 of the main text. 
As previously discussed in \ref{sec:master_equation}, a natural approximation consists in keeping only the $l_n=0$ terms in the sum, which leads to
\begin{equation}
    \delta \omega_n = -\frac{1}{2e} \sum_{l = -\infty}^{\infty} \tilde{I}^\KK( V + l \hbar \omega /e )  \matrixelement{n}{\hat{A}_{l} \hat{A}^\dagger_{l}}{n} \,. 
\end{equation}
We have dropped the mode index and the $A_l$ operator implicitly acts on the \unit{6}{\giga\hertz} mode. By using the identities $\matrixelement{n}{\hat{A}_{l} \hat{A}^\dagger_{l}}{n} = |\matrixelement{n}{\hat{A}_l}{n-l}|^2$ if $n \geq l \geq 0$ and $\matrixelement{n}{\hat{A}_{l} \hat{A}^\dagger_{l}}{n} = |\matrixelement{n-l}{\hat{A}_{-l}}{n}|^2$ if $l \leq 0$, we obtain
\begin{equation}
    \delta \omega_n = -\frac{1}{2e} \left[ \sum_{l = 1}^{n} \tilde{I}^\KK( V + l \hbar \omega /e ) \alpha_{n-l,l} +  \sum_{l = 0}^{\infty} \tilde{I}^\KK( V - l \hbar \omega /e ) \alpha_{n,l} \right]\,, 
\end{equation}
which corresponds to equation (3) of the main text. From the definition of the generalized Laguerre polynomials $L_0^{(l)}(\lambda^2)=1$ and $L_1^{(l)}(\lambda^2)=1+l-\lambda^2$, we arrive at the following expression for the shift of the $0 \rightarrow 1$ transition
\begin{equation}
    \delta \omega_{01} = -\frac{\lambda^2 e^{-\lambda^2}}{2e}  \sum_{l=0}^\infty \frac{\lambda^{2l}}{l!} \left[ \tilde{I}^{\KK}(V - (l+1) \hbar \omega /e ) + \tilde{I}^{\KK}(V - (l-1) \hbar \omega /e ) - 2 \tilde{I}^{\KK}(V - l \hbar \omega /e ) \right] \label{eq:LS2} \, ,
\end{equation}
which corresponds to equation (4) of the main text. Keeping only the $l=0$ term, we obtain a semi-classical expression, which is equivalent to the prediction of a classical model where the tunnel resistance is renormalized as $R_T \rightarrow R_T \Pi_n e^{\lambda_n^2}$.

\begin{figure}[h!]
    \centering
    \includegraphics{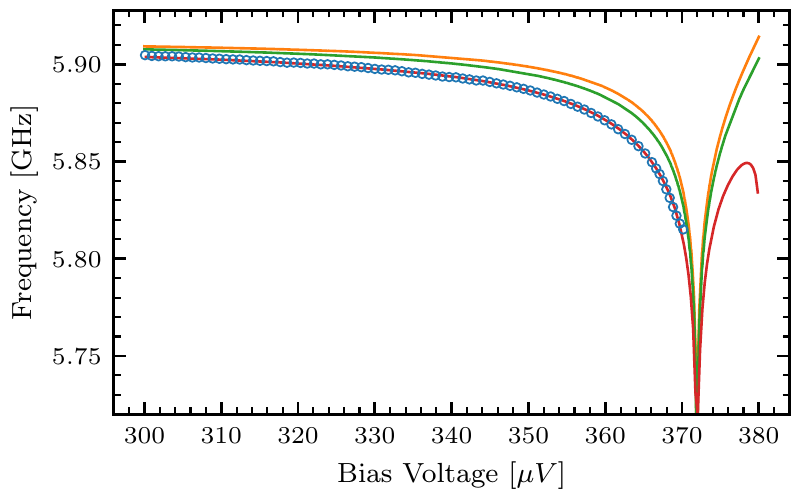}
    \caption{Comparison of the different predictions for the shift $\delta \omega_{01}$ of the fundamental resonance together with the experimental data (blue circles). The red curve corresponds to the multi-mode prediction (\ref{eq:LS1}), the green curve to the single mode approximation (\ref{eq:LS2}) and the orange curve to the semi-classical model, which is obtained by keeping only the $l=0$ term in (\ref{eq:LS2}).}
    \label{fig:LSapprox}
\end{figure}

We compare these different approximations and the exact expression in to our data in figure \ref{fig:LSapprox}. When we compare our data to the theoretical predictions, as here or as in figure 4 of the main text, we allow for a small frequency shift of the transition frequency (vertical offset) and also for a small shift of the gap (horizontal offset). With these adjustments, our data are in better agreement with the exact multi-mode calculation (\ref{eq:LS1}) than with the simplified formula obtained in (\ref{eq:LS2}). 

\bibliography{biblio}